\documentclass[12pt]{article}
\pdfoutput = 1
\usepackage{amsmath,amssymb,bm,epsfig,afterpage}
\usepackage{mathtools}
\usepackage{cite}
\usepackage{here}
\usepackage{color}
\usepackage{tabularx}
\usepackage{simplewick} 
\usepackage{bbm}
\usepackage[samesize]{cancel} 
\usepackage{comment}
\usepackage{nicematrix}
\usepackage{enumitem}

\renewcommand{\thefootnote}{\fnsymbol{footnote}}
\newcommand{\vev}[1]{{\langle{#1}\rangle}}

\newcommand{\br}[2]{\text{Br}({#1}\to{#2})}

\newcommand{\abs}[1]{\left|{#1}\right|}

\newcommand{\eps}{\epsilon}

\newcommand{\order}[1]{\mathcal{O}\left({#1}\right)}

\newcommand{\pr}{\prime}
\newcommand{\ppr}{{\prime\prime}}

\newcommand{\Ecal}{\mathcal{E}}

\newcommand{\Lcal}{\mathcal{L}}
\newcommand{\Mcal}{\mathcal{M}}

\newcommand{\Wcal}{\mathcal{W}}

\newcommand{\teps}{\tilde{\epsilon}} 
\newcommand{\heps}{\hat{\epsilon}}

\newcommand{\tD}{\widetilde{D}}

\newcommand{\eff}{{\mathrm{eff}}}

\newcommand{\MeV}{\mathrm{MeV}}
\newcommand{\GeV}{\mathrm{GeV}}
\newcommand{\TeV}{\mathrm{TeV}}

\newcommand{\CP}{{\mathrm{CP}}}

\newcommand{\VLL}{\mathrm{VLL}}
\newcommand{\VRR}{\mathrm{VRR}}
\newcommand{\LR}{\mathrm{LR}}
\newcommand{\SLL}{\mathrm{SLL}}
\newcommand{\SRR}{\mathrm{SRR}}

\newcommand{\bra}[1]{\langle{#1}|}
\newcommand{\ket}[1]{|{#1}\rangle}

\numberwithin{equation}{section}

\allowdisplaybreaks[3]
\newcolumntype{Y}{&gt;{\centering\arraybackslash}X} 

\usepackage{hyperref}

\definecolor{asparagus}{rgb}{0.53, 0.66, 0.42}
\definecolor{darkspringgreen}{rgb}{0.09, 0.45, 0.27}
\definecolor{darkturquoise}{rgb}{0.0, 0.81, 0.82}
\definecolor{dollarbill}{rgb}{0.52, 0.73, 0.4}

\begin{document}

\begin{titlepage}

\begin{flushright}
 {\tt
WU-HEP-26-01 
}
\end{flushright}

\vspace{1.2cm}
\begin{center}
{\Large
{\bf
Chiral enhancement in the vector-like fourth family: \\ 
Case of $b \to s \gamma$   
}
}
\vskip 2cm
Junichiro Kawamura$^{a}$~\footnote{junichiro-k@ruri.waseda.jp}
and 
Yuji Omura$^{b}$~\footnote{yomura@phys.kindai.ac.jp}  
\vskip 0.5cm

{\it $^a$
Department of Physics, Waseda University, Tokyo 169-8555, Japan.
}\\[3pt]

{\it $^b$
Department of Physics, Kindai University, Higashi-Osaka, Osaka 577-8502, Japan.  
}\\[3pt]

\vskip 1.5cm

\begin{abstract}
We demonstrate that a vector-like fourth family of quarks
induces a genuine chiral enhancement in $b\to s\gamma$,
which is absent in the Standard Model (SM).
The coexistence of doublet and singlet states allows the chirality flip 
to occur inside the loop, leading to contributions proportional to the heavy vector-like mass. 
The resulting amplitude is enhanced by a factor 
$\overline{\lambda}_d v_H/m_b$, 
which can be as large as $\mathcal{O}(40)$ for moderate Yukawa couplings. 
This leads to sizable deviation from the SM prediction 
even for $\mathcal{O}(\mathrm{TeV})$ vector-like quark masses and small mixing angles. 
We find that $\mathrm{Br}(\overline{B}\to X_s\gamma)$ provides the most stringent constraint 
on this scenario among a wide range of precision observables.
\end{abstract}
\end{center}
\end{titlepage}

\clearpage

\renewcommand{\thefootnote}{\arabic{footnote}}
\setcounter{footnote}{0}

\newcommand{\la}{{\lambda}}
\newcommand{\ka}{{\kappa}}
\newcommand{\mQ}{{m^2_{\tilde{Q}}}}
\newcommand{\mU}{{m^2_{\tilde{u}}}}
\newcommand{\mD}{{m^2_{\tilde{d}}}}
\newcommand{\mL}{{m^2_{\tilde{L}}}}
\newcommand{\mE}{{m^2_{\tilde{e}}}}
\newcommand{\mhu}{{m^2_{H_u}}}
\newcommand{\mhd}{{m^2_{H_d}}}
\newcommand{\ms}{{m^2_S}} 
\newcommand{\Ala}{{A_\lambda}}
\newcommand{\Aka}{{A_\kappa}} 

\newcommand{\id}[1]{\mathbf{1}_{#1}}
\newcommand{\ol}[1]{\overline{#1}}
\newcommand{\sg}{\sigma}
\newcommand{\sgb}{\overline{\sigma}}
\newcommand{\del}{\partial}

\newcommand{\htm}{\hat{m}}
\newcommand{\tU}{\widetilde{U}}
\newcommand{\SL}[2]{\mathrm{SL}({#1},{#2})}
\newcommand{\natN}{\mathbb{N}}
\newcommand{\intZ}{\mathbb{Z}}
\newcommand{\oGam}{\overline{\Gamma}} 
\newcommand{\Gam}{\Gamma} 
\newcommand{\Ita}{\mathrm{Im}\,\tau}

\newcommand{\QCD}{\mathrm{QCD}}
\newcommand{\CKM}{\mathrm{CKM}}
\newcommand{\SM}{\mathrm{SM}}
\newcommand{\hY}{\hat{Y}} 

\newcommand{\Acal}{\mathcal{A}}
\newcommand{\tM}{\tilde{M}}

\newcommand{\nop}[1]{\textcolor{blue}{#1}}
\newcommand{\LQCD}{\Lambda_{\mathrm{QCD}}}
\newcommand{\ds}[1]{\cancel{#1}}

\newcommand{\YA}[1]{{\color{darkviolet} [\textbf{YA:} #1]}}
\definecolor{darkviolet}{rgb}{0.58, 0.0, 0.83}
\newcommand{\violed}[1]{{\color{darkviolet}#1}}
\newcommand{\magenta}[1]{{\color{magenta}#1}}

\newcommand{\PQ}{\mathrm{PQ}}
\newcommand{\MSbar}{\overline{\mathrm{MS}}}

\newcommand{\dsl}{\ds{\ell}\,}
\newcommand{\dsa}{\ds{a}\,}
\newcommand{\dsb}{\ds{b}\,}
\newcommand{\dsc}{\ds{c}\,}
\newcommand{\dsp}{\ds{p}\,}
\newcommand{\dsq}{\ds{q}\,}
\newcommand{\dsk}{\ds{k}\,}
\newcommand{\dsP}{\ds{P}\,}
\newcommand{\dsQ}{\ds{Q}\,}

\newcommand{\dsdel}{\ds{\partial}\,}

\newcommand{\ox}{\overline{x}}
\newcommand{\oy}{\overline{y}}
\newcommand{\oz}{\overline{z}}

\newcommand{\ola}{\overline{\lambda}}
\newcommand{\oU}{\overline{U}}
\newcommand{\LcR}{L\leftrightarrow R} 
\newcommand{\Cb}{\overline{C}} 
\newcommand{\VLQ}{\mathrm{VLQ}} 
\newcommand{\tla}{\tilde{\lambda}}

\newcommand{\PLR}[2]{
\begin{pmatrix}
 {#1} \\ {#2} 
\end{pmatrix}
}

\tableofcontents

\newpage 

\section{Introduction}
\label{sec-Intro}

Flavor physics provides one of the most powerful probes of physics beyond the Standard Model (SM).
In the SM, flavor and CP violations are encoded 
in the Cabibbo-Kobayashi-Maskawa (CKM) matrix~\cite{Buras:1998raa}.  
Flavor physics is sensitive to new physics beyond the SM 
by looking at observables whose SM contributions
are suppressed by the small mixing angles of the CKM matrix. 
In particular, new physics contributions can easily dominate over the SM ones 
even if the new physics scale is much higher than the electroweak (EW) scale.

One of the most sensitive processes to new physics is $b\to s\gamma$.  
This process is represented by the dipole interaction 
\begin{align*}
 \Lcal_{\eff} = C_7 
 \frac{ie G_F}{8\sqrt{2}\pi^2} V_{ts}^* V_{tb} \; 
  m_b \left(\ol{s} \sigma^{\mu\nu} (1+\gamma_5) b\right)  F_{\mu\nu}, 
\end{align*}
with $\sigma^{\mu\nu}:= (1/2) [\gamma^\mu, \gamma^\nu]$. 
In the SM, 
the chirality flip in $b\to s\gamma$ occurs on the external $b$-quark line, 
and hence the amplitude is proportional to $m_b$.
The sensitivity becomes even stronger 
if the chirality flip is caused by a heavy fermion mass in the loop.
In the Minimal Supersymmetric SM (MSSM), for instance, 
the amplitude can receive contributions proportional to $m_{\tilde g}/m_b$, 
where $m_{\tilde{g}}$ is the gluino mass in the loop,       
leading to a parametrically enhanced contribution relative 
to the SM~\cite{Bobeth:1999ww}. 
This leads to some of the most stringent constraints 
in the MSSM parameter space~\cite{Gabbiani:1996hi}. 
Thus, the chirality flip of flavor changing dipole operators 
can provide a powerful probe of heavy new particles.

Among many possibilities of new particles, 
vector-like quarks are well motivated in new physics beyond the SM, 
such as models with grand unification~\cite{Gursey:1975ki,Achiman:1978vg,delAguila:1982fs,Hewett:1988xc,Barbieri:1994kw,Berezhiani:1995dt,Barger:1995dd,Iguro:2021kdw}, extra dimensions~\cite{Randall:1999ee},
and the solutions to the gauge hierarchy problem~\cite{Arkani-Hamed:2001nha,Arkani-Hamed:2002sdy,Perelstein:2003wd,Han:2003wu,Fajfer:2013wca}.  
There are also models introducing the vector-like fermions 
to understand the mass hierarchies of the SM fermions~\cite{Berezhiani:1983hm,Dimopoulos:1983rz,Berezhiani:1985in,Berezhiani:1991ds,Berezhiani:1992pj,Alonso:2018bcg,Smolkovic:2019jow,Arkani-Hamed:2026wwy}.  
The vector-like quarks have the rich phenomenology 
as discussed in e.g. Refs.~\cite{Branco:1986my,Higuchi:2000rh,Botella:2012ju,Alok:2015iha,Ishiwata:2015cga,Albergaria:2024pji}.   
The mixing of the vector-like quarks 
induces the flavor violations through the $Z$ boson~\cite{Langacker:1988ur,Nir:1990yq,Branco:1992wr,Barenboim:2001fd} 
and non-unitarity of the $3\times 3$ CKM matrix~\cite{Liao:2000re,Botella:2021uxz,Branco:2021vhs,Crivellin:2022rhw}. 
There are also studies on models with a $Z^\prime$ boson
in addition to the vector-like quarks~\cite{Altmannshofer:2014cfa,King:2017anf,Raby:2017igl,Darme:2018hqg,Kawamura:2019rth,Kawamura:2019hxp}.

We shall focus on the case in which vector-like quarks 
are embedded as a vector-like fourth family~\cite{delAguila:2000rc,Buras:2009ka,Arnan:2019uhr}.   
In this case, there are three types of vector-like quarks, 
namely $Q$, $U$ and $D$ whose gauge quantum numbers 
are respectively same as the $SU(2)_L$ doublet quark $q$ 
and the singlets $u$ and $d$. 
The coexistence of both doublet and singlet vector-like quarks 
allows for Yukawa interactions that generate chirality flips inside the loop, 
rather than the light external quark mass.  
As a result, 
the dipole amplitude receives contributions proportional to the heavy vector-like mass.
As explicitly derived in the main text, the effect is enhanced 
by the factor $\ol{\la}_d v_H/m_b \sim \ol{\la}_d \order{40}$, 
where $\ol{\la}_d$ is the Yukawa coupling constant  
and $v_H \sim 174~\GeV$ is the Higgs vacuum expectation value (VEV).  
This represents a genuine chiral enhancement that is absent in the SM and cannot be realized in models with only a single vector-like quark.

In this work, we investigate the chiral enhancement on $b\to s\gamma$ 
by the fourth family vector-like quarks.  
We shall derive the leading order (LO) formula 
for the Wilson coefficients relevant to $b\to s\gamma$.
Due to the mixing caused by the Yukawa coupling, 
the generalized CKM matrix is no longer unitary 
and the $Z$ and Higgs boson couplings violate flavor at the tree-level. 
We demonstrate that $b\to s\gamma$ provides the most stringent constraint on the parameter space, even surpassing tree-level flavor-changing neutral currents mediated by the $Z$ and Higgs bosons.

The rest of this paper is organized as follows. 
Our notations for the vector-like fourth family 
are introduced in Sec.~\ref{sec-model}.
The Wilson coefficients relevant to $b\to s\gamma$ 
is derived in Sec.~\ref{sec-calcW}. 
Section~\ref{sec-pheno} is devoted 
to the phenomenology of the vector-like fourth family quarks, 
with particular interest in the chiral enhancement effect. 
Numerical analysis is in Sec.~\ref{sec-analysis}.  
Section~\ref{sec-concl} concludes this paper.  
In App.~\ref{app-analytical}, 
we show the approximate formula of the couplings and the Wilson coefficients.
The numerical coefficients used in our analysis are derived in App.~\ref{app-coes}.
We derive 1-loop formula of $\gamma\mathrm{-}Z$ mixing in App.~\ref{app-AZmixing} 
to show the gauge invariance of the total amplitude 
without relying on the unitarity of the CKM matrix.

\section{Masses and interactions of vector-like fourth family}
\label{sec-model}

We consider the extension of the SM only by the fourth family vector-like quarks, 
\begin{align}
Q = 
\begin{pmatrix}
 Q_u \\ Q_d
\end{pmatrix}, 
\quad 
U, \quad D, 
\end{align}
whose gauge quantum numbers are the same 
as the SM quarks, $q = (q_u, q_d)$, $u$ and $d$, respectively. 
In addition, there are the vector-like counterparts 
$\ol{Q}$, $\ol{U}$ and $\ol{D}$, respectively, 
so that these have vector-like masses without the Higgs mechanism. 
The vector-like quark masses and the Yukawa couplings are given by 
\begin{align}
\label{eq-LNP}
 -\Lcal =&\ m_Q \ol{Q} P_L Q + m_U \ol{U} P_L U + m_D\ol{D} P_L D  \\ \notag 
         &\ + y^u_{ij} \ol{u}_i P_L q_j \cdot H
            - y^d_{ij} \ol{d}_i P_L q_j \cdot \tilde{H}               \\ \notag 
         &\ + \eps^u_{L_j} \ol{U}    P_L q_j \cdot H  
            + \eps^u_{R_i} \ol{u}_i  P_L Q   \cdot H  
            - \eps^d_{L_j} \ol{D}    P_L q_j  \cdot\tilde{H}
            - \eps^d_{R_i} \ol{d}_i  P_L Q    \cdot\tilde{H}           \\ \notag 
         &\ + \la_u \ol{U} P_L Q \cdot H 
            - \la_d \ol{D} P_L Q  \cdot\tilde{H}  
            + \ola_u \ol{Q} P_L U \tilde{H} + \ola_d \ol{Q} P_L D H + h.c., 
\end{align}
where $\tilde{H} := \epsilon  H^*$ 
with $\epsilon=i\sigma_2$, and $H=(H^+, H^0)$.    
Here, $i,j=1,2,3$ run over the three flavors of the SM quarks. 
The chiral projections are defined 
as $P_L := (1-\gamma_5)/2$ and $P_R := (1+\gamma_5)/2$. 
The multiplication with $\cdot$ is for the $SU(2)_L$ doublets 
with $\epsilon$, whereas that without it is with identity, 
e.g. $Q\cdot H = Q_u H^0 - Q_d H^+$ and 
$\ol{Q}H = \ol{Q}_u H^+ + \ol{Q}_d H^0$.
The first line is the vector-like masses, 
and the second line is the SM Yukawa couplings. 
The couplings between the SM flavors and the fourth family are in the third line. 
The last line is the Yukawa couplings in the fourth family, 
which appear only if both of $Q$ and $(U,D)$ are introduced. 
Without changing the gauge interactions, 
the vector-like mass terms $\ol{Q} P_L q_j, \ol{u}_i P_L U$ and $\ol{d}_i P_L D$ 
can be rotated away.  
In addition, either of the SM Yukawa coupling matrices can be diagonalized, 
and we start from the basis, $y^u_{ij} = y^u_i \delta_{ij}$.   
Throughout this work, we start with this basis without loss of generality.

After the EW symmetry breaking, 
the quark mass terms are given by 
\begin{align}
 -\Lcal_\mathrm{mass} =&\ 
 \ol{u}_R \Mcal_u u_L + \ol{d}_R \Mcal_d d_L  + h.c. 
\\ \notag 
:=&\ 
\ol{u}_R 
\begin{pmatrix}
 y^u_{i}\delta_{ij} v_H      & \eps^u_{R_i} v_H & 0_i \\
 \eps^u_{L_j} v_H  & \la_u v_H & m_U \\ 
 0_j & m_Q & \ola_u v_H
\end{pmatrix}
u_L  
+ 
\ol{d}_R 
\begin{pmatrix}
 y^d_{ij} v_H      & \eps^d_{R_i} v_H & 0_i \\
 \eps^d_{L_j} v_H  & \la_d v_H & m_D \\ 
 0_j & m_Q & \ola_d v_H
\end{pmatrix}
d_L + h.c., 
\end{align}
where 
\begin{align}
\label{eq-gBasis}
u_L := 
P_L
\begin{pmatrix}
 q_{u,i} \\ Q_u \\ U 
\end{pmatrix}, 
\quad 
u_R := 
P_R 
\begin{pmatrix}
 u_i \\ U \\ Q_u 
\end{pmatrix}, 
\quad  
d_L :=  
P_L 
\begin{pmatrix}
 q_{d,i} \\ Q_d \\ D
\end{pmatrix},
 \quad 
 d_R := P_R
\begin{pmatrix}
 d_i \\ D \\ Q_d 
\end{pmatrix}. 
\end{align}
The mass matrices are diagonalized as 
\begin{align}
 U_{u_R}^\dag \Mcal_u U_{u_L} 
=&\ D_u := \mathrm{diag}\left(m_u, m_c, m_t, m_{U_4}, m_{U_5}\right), 
\\ \notag 
 U_{d_R}^\dag \Mcal_d U_{d_L} 
=&\ D_d := \mathrm{diag}\left(m_d, m_s, m_b, m_{D_4}, m_{D_5}\right),  
\end{align}
where $U_{q_X}$, with $q=u,d$ and $X=L,R$, is a $5\times 5$ unitary matrix.

In the mass basis, the Dirac fermions are defined as 
\begin{align}
 \psi_u := U_{u_L}^\dag u_L + U_{u_R}^\dag u_R, 
\quad 
 \psi_d := U_{d_L}^\dag d_L + U_{d_R}^\dag d_R. 
\end{align}
We denote the gauge couplings as 
\begin{align}
 \Lcal_g = \sum_{X=L,R} 
           \left( W^+_\mu \ol{\psi}_u \gamma^\mu g^W_{X} P_X \psi_d + h.c. 
              +   \sum_{q=u,d} Z_\mu \ol{\psi}_q \gamma^\mu g^Z_{q_X} P_X \psi_q 
          \right), 
\end{align}
where the coupling matrices are given by 
\begin{align}
g^W_{X} = \frac{g}{\sqrt{2}} U_{u_X}^\dag H_X U_{d_X}, 
\quad 
g^Z_{q_X} 
= \frac{g}{c_W} \left( I_q\, U_{q_X}^\dag H_X U_{q_X} - s_W^2 Q_q \id{5}\right). 
\end{align}
The iso-spin and electric charge of the quarks 
are normalized as $(Q_u, I_u) = (2/3, 1/2)$ and $(Q_d, I_d) = (-1/3, -1/2)$. 
Here, $c_W = \cos \theta_W$ ($s_W=\sin\theta_W$) 
is the cosine (sine) of the weak mixing angle.  
The projection matrices 
\begin{align}
 H_L = \mathrm{diag}(1,1,1,1,0), 
\quad 
 H_R = \mathrm{diag}(0,0,0,0,1),   
\end{align} 
are given in the gauge basis Eq.~\eqref{eq-gBasis}. 
We note that the extended CKM matrix 
$U^\dag_{u_L} H_L U_{d_L}$ is not unitary 
due to the mixing between the doublets and the singlets.

We decompose the Higgs doublet as 
\begin{align}
 H = 
\begin{pmatrix}
 0 \\ v_H
\end{pmatrix}
+ \frac{1}{\sqrt{2}}
\begin{pmatrix}
 - \sqrt{2} ia_+ \\ h + i a_0
\end{pmatrix},
\end{align}
where $a_{0}$, $a_+$ and $a_- := a_+^*$ 
are the Nambu-Goldstone (NG) bosons of the EW gauge symmetry. 
Here the phase of the charged mode is chosen 
so that the Yukawa couplings have formally the same form in Eq.~\eqref{eq-YNGmodes}. 
The Yukawa couplings to the Higgs boson $h$ are given by  
\begin{align}
- \Lcal_H = h \sum_{q=u,d} \ol{\psi}_q Y_q^h P_L \psi_q + h.c.,    
\end{align}
where 
\begin{align}
 Y_q^h := \frac{1}{\sqrt{2}} U_{q_R}^\dag  
\begin{pmatrix}
 y^q_{ij}   & \eps^q_{R_i} & 0_i        \\ 
\eps^q_{L_j}& \la_q        & 0          \\ 
        0_j &            0 & \ol{\la}_q \\ 
\end{pmatrix}
 U_{q_L},  
\end{align}
for $q=u,d$. 
The Yukawa couplings to the NG modes,
\begin{align}
 -\Lcal_a = a_+ \ol{\psi}_u Y^{a}_+ P_L \psi_d + 
            a_- \ol{\psi}_d Y^{a}_- P_L \psi_u +  
            a_0 \sum_{q=u,d} \ol{\psi}_q Y^{a}_q P_L \psi_q + h.c.,   
\end{align} 
are expressed by the gauge couplings as 
 \begin{align}
\label{eq-YNGmodes}
Y^{a}_q =&\ 
     \frac{i}{m_Z} \left(D_q g^Z_{q_L} - g^Z_{q_R} D_q \right), 
\\ \notag
Y^{a}_+ =&\ 
     \frac{i}{m_W} \left(D_u g^W_{L} - g^W_{R} D_d \right), 
\quad 
Y^{a}_- =
     \frac{i}{m_W} \left\{D_d \left(g^W_{L}\right)^\dag 
                           - \left(g^W_{R}\right)^\dag  D_u \right\}.  
\end{align}
We note that 
the $Z$ and Higgs bosons also have flavor violating couplings, 
induced by the mixing between the SM flavors and the vector-like quarks.

\section{Calculation of Wilson coefficients}
\label{sec-calcW}

In the SM, the relevant effective interactions 
for $b \to s \gamma$ are given by\cite{Buras:1998raa},    
\begin{align}
\label{eq-Leff}
 \Lcal_{\eff} = \frac{G_F}{\sqrt{2}} V_{ts}^* V_{tb}  
 \sum_{i=1}^8 C_i(\mu_b) Q_i,   
\end{align}
where $V_{tq}^* V_{tb}$ is the CKM matrix in the SM. 
The operators are defined as 
\begin{align}
\label{eq-Q18}
 Q_1 =&\ \left(\ol{c}_\alpha b_\beta\right)_{V-A}
         \left(\ol{s}_\beta  c_\alpha\right)_{V-A}, 
\quad &  
 Q_2 =&\ \left(\ol{c} b\right)_{V-A}
         \left(\ol{s} c\right)_{V-A}, 
\\ \notag 
 Q_3 =&\  \left(\ol{s}b\right)_{V-A}
          \sum_{q} \left(\ol{q} q\right)_{V-A}, 
\quad &
 Q_4 =&\  \left(\ol{s}_\alpha b_\beta\right)_{V-A}
          \sum_q \left(\ol{q}_\beta  q_\alpha\right)_{V-A}, 
\\ \notag 
 Q_5 =&\  \left(\ol{s}b\right)_{V-A}
          \sum_{q} \left(\ol{q} q\right)_{V+A}, 
\quad &
 Q_6 =&\  \left(\ol{s}_\alpha b_\beta\right)_{V-A}
          \sum_q \left(\ol{q}_\beta  q_\alpha\right)_{V+A}, 
\\ \notag 
 Q_7 =&\  \frac{ie}{8\pi^2}m_b 
          \left(\ol{s} \sigma^{\mu\nu} (1+\gamma_5) b \right) F_{\mu\nu}, 
\quad &
 Q_8 =&\  \frac{ig_s}{8\pi^2}m_b  
          \left(\ol{s} \sigma^{\mu\nu} T^a (1+\gamma_5) b \right) G^a_{\mu\nu}, 
\end{align}
where $\alpha,\beta$ are the color indices 
and $q$ runs over the quarks. 
Here, $\sigma^{\mu\nu}:= (1/2) [\gamma^\mu, \gamma^\nu]$.
The fermion bi-linears  are defined as  
$(\ol{q}_1 q_2)_{V\pm A} = \ol{q}_1 \gamma_\mu (1\pm \gamma_5) q_2$.

In our case, there are right-current charged interactions,   
which induce the left-right mixing operators 
\begin{align}
\label{eq-Q910}
 Q_9 := \frac{m_b}{m_c} \left( \ol{c}_\alpha b_\beta \right)_{V-A}
        \left( \ol{s}_\beta c_\alpha \right)_{V+A}, 
\quad
 Q_{10} := \frac{m_b}{m_c} \left( \ol{c}_\alpha b_\alpha \right)_{V-A}
        \left( \ol{s}_\beta c_\beta \right)_{V+A}, 
\end{align}
analogous to $Q_1$ and $Q_2$ respectively, but with the mass ratio factor. 
In addition, there are also primed operators which can be obtained 
by flipping the chirality $\LcR$ of $Q_{1\mathrm{-}10}$.

The mixing with the fourth family 
also induce the flavor violating neutral interactions~\cite{Buras:2011zb}, 
\begin{align}
\label{eq-Qnn}
 \Lcal_{\eff}^{nn} = \frac{G_F}{\sqrt{2}} V_{ts}^* V_{tb} 
                 \sum_{q} \sum_{j=1,2} \sum_{X,Y=L,R}  
                 \left[ 
                  C_a^q(X,Y) Q_a^q(X,Y) 
     +\delta_{qd}\hat{C}_a^d(X,Y) \hat{Q}_a^d(X,Y)  
                 \right]  
\end{align}  
In general, there are 48 operators~\footnote{
Some of the operators are included in $Q_{4-6}$. 
Here, we treat these new operators to be induced by the flavor-violating 
$Z$ boson couplings, 
while the conventional ones are from the penguin diagrams 
which are subdominant for $b\to s\gamma$. 
} 
\begin{align}
Q_{1}^{q}(X,Y) := \left(\ol{s}_\alpha b_\beta \right)_{X}  
                  \left(\ol{q}_\beta q_{\alpha} \right)_{Y}, 
\quad    
Q_{2}^{q}(X,Y) := \left(\ol{s}_\alpha b_\alpha \right)_{X}  
                  \left(\ol{q}_\beta q_{\beta} \right)_{Y}, 
\end{align}
and additional 8 operators for $q=d$
\begin{align}
\hat{Q}_{1}^{d}(X,Y) := \left(\ol{s}_\alpha d_\beta \right)_{X}  
                        \left(\ol{d}_\beta  b_{\alpha} \right)_{Y}, 
\quad    
\hat{Q}_{2}^{d}(X,Y) := \left(\ol{s}_\alpha d_\alpha \right)_{X}  
                            \left(\ol{d}_\beta  b_{\beta} \right)_{Y},  
\end{align}
where $X,Y = L,R$.  
Here, we denote  
$(\ol{s}_\alpha d_\alpha )_{L} := (\ol{s}_\alpha d_\alpha )_{V-A}$
and it is the same for the other quark bi-linears.    
We neglect the scalar operators induced by the Higgs boson, 
since their Wilson coefficients are suppressed by $m_q^2/m_W^2$, 
with $m_q$ being quark mass, compared with the vector operators.

The dominant contribution to $b\to s\gamma$ comes from $C_7$ in the SM, 
and sub-dominantly from $C_2$ and $C_8$ through the operator mixing 
due to the renormalization group (RG) evolution. 
We shall calculate the 1-loop contribution to $C_7^{(\pr)}$ and $C_8^{(\pr)}$ 
of the fourth family quark model in the rest of this section.    
All the other operators are calculated at the tree-level.

\subsection{Amplitudes of the diagrams}

\begin{figure}[th]
  \centering
  \includegraphics[width=0.90\linewidth]{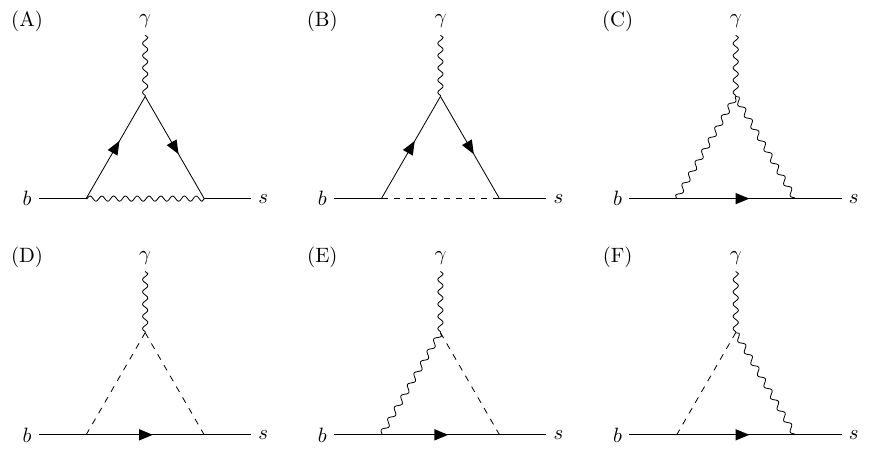}
  \caption{\label{fig-FDs}
Diagrams contribute to the Wilson coefficient $C_7$.
Those for $C_8$ can be obtained by replacing 
the photon $\gamma$ to the gluon $g$ in the diagrams (A) and (B). 
}
\end{figure}

We calculate the amplitudes relevant to $b\to s\gamma$. 
The Feynman diagrams (A)-(F) are shown in Fig.~\ref{fig-FDs},
where the solid lines are the quarks $f=\psi_{q}$, 
the wavy lines are the vector bosons $V=W_\pm, Z$ 
and 
the dashed lines are the scalars $S=h, a_0, a_\pm$.

In our calculation, we work on the Feynman-'t Hooft gauge and 
neglect $\order{m_b^2}$ contributions which are sub-leading.  
The amplitudes of the diagrams (A)-(F) in Fig.~\ref{fig-FDs}
are calculated as 
\begin{align}
\Mcal_A^\mu =&\ \frac{eQ_q}{16\pi^2m_V^2} \ol{s} \Bigl[
\gamma^\mu  \Bigl\{
 m_V^2 \left(\frac{1}{\eta_V} -2 - 2C_{00} \right)
 + \left( C_0+C_{p}+C_{k}+C_{pk} \right) q^2  
+ m_a^2 C_0 
\Bigr\}   g_{La}^\dag g_{La} 
\\ \notag 
&\ \hspace{1.5cm} 
+ 2 p^\mu \Bigl\{
  m_s (C_0+C_p+C_k+C_{kp}) g_{La}^* g_{La}  
+ m_b (C_{pp}+C_p)  g_{Ra}^\dag g_{Ra} 
\\ \notag 
&\ \hspace{3.5cm} 
- m_a (C_0+2C_p) g_{Ra}^\dag g_{La}  
\Bigr\} 
\\ \notag 
&\ \hspace{1.5cm} 
+ 2k^\mu \Bigl\{
   m_s (C_{kk}+C_k) g_{La}^\dag g_{La}     
 + m_b (C_0+C_p+C_k+C_{kp}) g_{Ra}^\dag g_{Ra} 
\\ \notag  
&\ \hspace{3.5cm} 
 -m_a (C_0+2C_k) g_{Ra}^\dag g_{La} 
\Bigr\} 
\Bigr] 
P_L b + (L\leftrightarrow R), 
\\ 
\Mcal_B^\mu =&\ \frac{eQ_q}{32\pi^2m_S^2} \ol{s} \Bigl[
\gamma^\mu \Bigl\{
\Bigl(
m_S^2\left(\frac{1}{\eta_S}-1-2C_{00}\right) + C_{pk} q^2  + m_a^2 C_0 
\Bigr) y_{La}^\dag y_{La} 
 \notag \\
&\ \hspace{3.5cm}
+m_a C_0  \left(m_b y_{La}^\dag y_{Ra} + m_s y_{Ra}^\dag y_{La} \right)
\\ \notag 
&\ \hspace{1.5cm}
+ 2p^\mu \Bigl\{ m_a C_p y_{Ra}^\dag y_{La} 
+ m_b (C_{pp}+C_p) y_{Ra}^\dag y_{Ra} + m_s C_{pk} y_{La}^\dag y_{La}  
\Bigr\}
\\ \notag 
&\ \hspace{1.5cm}
+2k^\mu \Bigl\{
  m_a C_k y_{Ra}^\dag y_{La} 
+ m_b C_{pk} y_{Ra}^\dag y_{Ra} + m_s (C_{kk}+C_k) y_{La}^\dag y_{La} 
\Bigr\}
\Bigr] P_L b + (L\leftrightarrow R), 
\\ 
\Mcal_C^\mu =&\ -\frac{eQ_W}{32\pi^2m^2_W} \ol{s} \Bigl[
 \gamma^\mu \Bigl\{
\left(
 6m_W^2 \left(\frac{1}{\eta_W}-\frac{2}{3} - 2\Cb_{00}\right) 
 +2(\Cb_{pk}+\Cb_p+\Cb_k) q^2\right) g_{La}^\dag g_{La} 
\\ \notag 
&\ \hspace{3.5cm}
 - 3m_a \Cb_0 \left(m_b g_{La}^\dag g_{Ra} + m_s g_{Ra}^\dag g_{La} \right) 
\Bigr\} 
\\ \notag 
&\ \hspace{0.5cm}
- 2p^\mu \Bigl\{
m_s (2\Cb_{kp}-\Cb_p-\Cb_k) g_{La}^\dag g_{La} 
    + m_b (2\Cb_{pp}+\Cb_p) g_{Ra}^\dag g_{Ra}
  + 3m_a \Cb_p g_{Ra}^\dag g_{La} 
\Bigr\}
\\ \notag 
&\ \hspace{0.5cm}
- 2k^\mu \Bigl\{
    m_s (2\Cb_{kk}+\Cb_k) g_{La}^\dag g_{La} 
  + m_b (2\Cb_{pk}-\Cb_p-\Cb_k) g_{Ra}^\dag g_{Ra} 
  + 3m_a \Cb_k g_{Ra}^\dag g_{La} 
\Bigr\}
\Bigr] P_L b 
\\ \notag 
&\ \hspace{0.5cm} + (L\leftrightarrow R), 
\\ 
\Mcal^\mu_D =&\ \frac{eQ_W}{32\pi^2m_W^2} \ol{s} \Bigl[
 -m_W^2 \gamma^\mu \left(\frac{1}{\eta_W} - 2\Cb_{00}\right)  
 y_{La}^\dag y_{La} 
\\ \notag 
&\ \hspace{0.1cm} 
 + p^\mu \left\{
   m_s(2\Cb_{pk}+\Cb_k) y_{La}^\dag y_{La}  
 + m_b (2\Cb_{pp}+\Cb_p) y_{Ra}^\dag y_{Ra} 
 - m_a (2\Cb_p + \Cb_0) y_{Ra}^\dag y_{La}  
\right\}
\\ \notag 
&\ \hspace{0.1cm} 
 + k^\mu \left\{
   m_s (2\Cb_{kk}+\Cb_k) y_{La}^\dag y_{La} 
 + m_b (2\Cb_{kp}+\Cb_p) y_{Ra}^\dag y_{Ra} 
 - m_a (2\Cb_k + \Cb_0)  y_{Ra}^\dag y_{La} 
\right\}
\Bigr] P_L b 
\\ \notag 
&\ \hspace{0.1cm} 
+ \left(L\leftrightarrow R\right), 
\\ 
\Mcal_{E}^\mu =&\ \frac{ieQ_W}{32\pi^2m_W} \ol{s} \left[
 \gamma^\mu \left\{ 
  m_s \Cb_k y_{Ra}^\dag g_{La} 
 -m_b \Cb_p y_{La}^\dag g_{Ra} 
 -m_a \Cb_0 y_{La}^\dag g_{La} 
 \right\}
 + 2p^\mu \Cb_p y_{Ra}^\dag g_{La} 
\right] P_L b 
\notag  \\ 
&\ 
\hspace{0.1cm}
+ \left(L\leftrightarrow R\right), 
\\ 
\Mcal_{F}^\mu =&\ -\frac{ieQ_W}{32\pi^2m_W} \ol{s} \left[
 \gamma^\mu \left\{ 
  - m_s \Cb_k g_{Ra}^\dag y_{La} 
  + m_b \Cb_p g_{La}^\dag y_{Ra} 
  - m_a \Cb_0 g_{La}^\dag y_{La} 
 \right\}
 + 2 k^\mu \Cb_k  g_{Ra}^\dag y_{La} 
\right] P_L b 
\notag  \\ 
&\ \hspace{0.1cm}
+ \left(L\leftrightarrow R\right),  
\end{align}
where $p^\mu$ is an incoming momentum of $b$ quark
and $k^\mu$ is an outgoing momentum of $s$ quark. 
The outgoing photon momentum is $q^\mu = p^\mu-k^\mu$. 
Here, $Q_W = +1$ is the electric charge of the $W$ boson 
and $Q_q$ is that of the internal quarks, 
with $Q_u = 2/3$ and $Q_d = -1/3$. 
In the diagrams (A) and (B), 
the internal quark is up-type (down-type) quarks  
in the $W/a_\pm$ ($Z/a/h$) boson loop, 
whereas there is only the $W$ boson loop for the other diagrams.

For the $W$ boson loops, the gauge couplings are given by 
\begin{align}
 g_{Xa}      = \left[g^W_{X} \right]_{a3}, 
\quad  
 g_{Xa}^\dag = \left[g^W_{X} \right]^*_{a2},   
\end{align}
and those for the $Z$ boson loops are obtained 
by formally replacing $g^W_X \to g^Z_{d_X}$. 
The Yukawa couplings to the charged NG bosons are given by 
\begin{align}
\label{eq-yLRa}
 y_{La} = \left[Y_+^a \right]_{a3}, \quad 
 y_{Ra} = \left[Y_-^a \right]_{3a}^*, \quad 
 y_{La}^\dag = \left[Y_+^a \right]^*_{a2}, \quad 
 y_{Ra}^\dag = \left[Y_-^a \right]_{2a}. 
\end{align}
Those for the neutral scalars $S = a, h$ can be obtained 
by replacing $Y_{\pm}^a \to Y^S_d$.

Throughout this work, 
we use the dimensional regularization in $n=4-2\eps$ dimensional spacetime. 
The loop functions in (A) and (B) are defined as  
\begin{align}
 \int \frac{d^n \ell}{i(2\pi)^n} \frac{1}{D_3} 
 =: &\ \frac{1}{32\pi^2m_B^2} C_0, 
\\ \notag  
 \int \frac{d^n \ell}{i(2\pi)^n} \frac{\ell^\mu}{D_3} 
 =:&\ \frac{1}{32\pi^2m_B^2} \left(p^\mu C_p + k^\mu C_k  \right), 
\\ \notag  
 \int \frac{d^n \ell}{i(2\pi)^n} \frac{\ell^\mu\ell^\nu} {D_3} 
=: &\ \frac{1}{32\pi^2m_B^2} 
  \left[m_B^2 g^{\mu\nu}\left(-\frac{1}{2\eta_B} + C_{00} \right)  
 + \sum_{(p_1,p_2)= \left\{p,k\right\}} p_1^\mu p_2^\nu C_{p_1p_2} 
 \right],
\end{align}
where the denominator is 
\begin{align}
D_3 := (m_B^2-\ell^2)\left\{m_a^2-(\ell+p)^2\right\} 
                     \left\{m_a^2-(\ell+k)^2\right\},  
\end{align}
with $m_B$ being the mass of the internal boson $B=W,Z,a_\pm,a,h$
and $m_a$ being that of the internal fermion. 
For the diagrams (C)-(F), 
the loop functions $\Cb_l$, with $l=0,p,k,00,pp,pk,kk$, 
are defined in the same way but the denominator in the loop integration 
is replaced by    
\begin{align}
\ol{D}_3 := (m_a^2-\ell^2)\left\{m_W^2-(\ell+p)^2\right\} 
                          \left\{m_W^2-(\ell+k)^2\right\}.  
\end{align}
The divergent factor is defined as 
\begin{align}
 \frac{1}{\eta_B} := \frac{1}{\eps} -\gamma + \log \frac{4\pi}{m_B^2}. 
\end{align}

In addition to the vertex corrections, 
the self-energy corrections to the quarks should also be added. 
The amplitudes from the loops mediated by the vector boson $V$ and scalar $S$ 
are respectively given by 
\begin{align}
 \Mcal_{V}^\mu =&\ -\frac{eQ_d}{16\pi^2 } \ol{s} \gamma^\mu
 \Biggl[\left(\frac{1}{\eta_V} - 1 + B_1(0) \right) g_{La}^\dag g_{La} 
\\ \notag 
 &\ \hspace{4.0cm}  
-4m_a B^\pr_{0}(0) \left(m_b g_{La}^\dag g_{Ra} + m_s g_{Ra}^\dag g_{La} \right) 
 \Biggr] P_L b 
+ \left(L\leftrightarrow R\right), 
\\ \notag  
 \Mcal_{S}^\mu =&\ -\frac{eQ_d}{32\pi^2 } \ol{s} \gamma^\mu
 \Biggl[ \left(\frac{1}{\eta_S}  + B_1(0) \right) y_{La}^\dag y_{La} 
\\ \notag 
&\ \hspace{4.0cm}
   +2m_a B^\pr_{0}(0) \left(m_b y_{La}^\dag y_{Ra} + m_s y_{Ra}^\dag y_{La} \right) 
 \Biggr] P_L b 
+ \left(L\leftrightarrow R \right), 
\end{align}
where the two-point loop functions are defined as 
\begin{align}
 \int \frac{d^n\ell}{i(2\pi)^n} 
    \frac{1}{D_2(p)}  
 =:&\ \frac{1}{16\pi^2} \left(\frac{1}{\eta_B} + B_0(p^2)\right),  
\\ \notag 
 \int \frac{d^n\ell}{i(2\pi)^n} 
    \frac{\ell^\mu}{D_2(p)} 
 =:&\ -\frac{p^\mu}{32\pi^2} \left(\frac{1}{\eta_B} + B_1(p^2)\right), 
\end{align}
and 
\begin{align}
 B^\pr_1(0) := \left. \frac{d B_1(p^2)}{d p^2} \right|_{p^2=0}. 
\end{align}
Here, the denominator is given by 
\begin{align}
 D_2(p) := \left(m_a^2-\ell^2\right)\left\{m_B^2 -(\ell+p)^2\right\}. 
\end{align}

\subsection{Amplitudes of \texorpdfstring{$W$}{W} boson loops}

After adding the diagrams (A)-(F) 
and the self-energy corrections to the quarks, 
both vector boson and scalar loops, 
the amplitude involving the $W$ boson and its NG mode reads 
\begin{align}
\label{eq-MaW}
 \Mcal_{\gamma W}^\mu =&\ \frac{eG_F}{4\sqrt{2}\pi^2} \ol{s} \sum_{a} \Bigl[
- \gamma^\mu \left(\frac{2m_W^2}{\eta_W} + \frac{q^2}{3} \right)   \la^W_{aLL}
   + (q^2\gamma^\mu - \dsq q^\mu) D_W(r_{aW}) \la^W_{aLL}
\\ \notag 
&\hspace{1.5cm}
+ \sigma^{\mu\nu}q_\nu \Bigl\{ 
 D_W^\pr (r_{aW}) \left(m_b \la^W_{aRR} + m_s \la^W_{aLL} \right) 
+ m_a D^\ppr_W(r_{aW}) \la^W_{aRL}  
\Bigr\}
\Bigr] P_L b + \left(L\leftrightarrow R\right), 
\end{align} 
where $r_{aB} := m_{U_a}^2/m_W^2$ and 
$\la_{aXY}^W := (2/g^2)\left[g^W_X \right]^{*}_{a2} \left[g^W_Y\right]_{a3}$. 
The latter is analogous to the usual one, e.g. $\la_t := V_{ts}^* V_{tb}$. 
The loop functions are defined as 
\begin{align}
 D_W(r) 
:= &\ \frac{26}{27}  
-\frac{4}{9} \log r - \frac{r^2(25-19r)}{36(1-r)^3} 
                    + \frac{r^2(-6-2r+5r^2)\log r}{18(1-r)^4},
\\ \notag 
 D_W^\pr(r) :=&\  
-\frac{23}{18} 
+ \frac{r(7-5r-8r^2)}{12(1-r)^3} + \frac{r^2(2-3r)\log r}{2(1-r)^4},  
\\ \notag 
 D_W^\ppr(r):=&\
 \frac{10}{3} + 
 \frac{r(3-5r)}{2(1-r)^2} + \frac{r(2-3r)\log r}{(1-r)^3}.  
\end{align}
The functions $D_W$ and $D_W^\pr$ respectively match 
to the Inami-Lim functions $D_0$ and $D_0^{\pr}$ up to the constants. 
We keep the constant terms 
because the GIM mechenism does not work exactly in our case. 
The contribution with $m_a D_W^{\pr\pr}$ 
includes the chirally enhanced effect due to the vector-like fourth family quarks,
which is the main scope of this paper~\footnote{
The form of $D_W^{\pr\pr}$ is 
consistent with Eq.~(3.2) of Ref.~\cite{Cho:1993zb}, 
studying the $W_R$ contributions in the LR model up to normalization.}.

We note that the first term in Eq.~\eqref{eq-MaW} is not vanishing 
and hence it is divergent and violates the gauge symmetry 
since the GIM mechanism does not work exactly, i.e. $\sum_a \la_{aLL}^W \ne 0$.  
Using the result of App.~\ref{app-AZmixing}, 
this contribution is canceled by the 1-loop effects 
to the photon mixing with the $Z$ boson and its NG mode. 
The amplitude due to the mixing is given by 
 \begin{align}
\label{eq-dltMbsg}
 \delta \Mcal^\mu_{\gamma Z}    
+
 \delta \Mcal^\mu_{\gamma a_0}    
=&\ 
 \frac{eG_F} {4\sqrt{2}\pi^2} \ol{s}  \Bigl[
\gamma^{\mu}  \left(\frac{2m_W^2}{\eta_W} + \frac{q^2}{3}\right) 
\\ \notag 
&\hspace{0.2cm}
+ \left\{\left(\frac{1}{6}+5c_W^2
 \right)\frac{1}{\eta_W} 
+ \frac{2}{3} c_W^2 \right\}
 \left(q^2 \gamma^{\mu} - q^\mu \dsq\right)
\Bigr] 
\left(\sum_{a} \la^W_{aLL}\right)  P_L b  
+ \left(L\leftrightarrow R\right).   
\end{align}  
Thus, the gauge violating part in Eq.~\eqref{eq-MaW} is canceled. 
The contribution in the second line would be absorbed
by the renormalization of the neutral SM gauge bosons 
which is subdominant for $b\to s\gamma$.

\subsection{Amplitudes of \texorpdfstring{$Z/H$}{Z/H} boson loops}

There are also contributions from the loops mediated by 
the $Z$ and Higgs bosons which, in general, violate flavor.   
For $B = Z, H$, we find 
\begin{align}
\label{eq-MaB}
 \Mcal_{\gamma B}^\mu =&\ \frac{eQ_d G_F}{4\sqrt{2}\pi^2} \ol{s} \sum_{a} \Bigl[
   (q^2\gamma^\mu - \dsq q^\mu) E_B(r_{aB}) \la^B_{aLL}
\\ \notag 
&\hspace{1.5cm}
+ \sigma^{\mu\nu}q_\nu \Bigl\{ 
 E_B^\pr (r_{aB}) \left(m_b \la^B_{aRR} + m_s \la^B_{aLL} \right) 
+ m_a E^\ppr_B(r_{aB}) \la^B_{aRL}  
\Bigr\}
\Bigr] P_L b + \left(L\leftrightarrow R\right), 
\end{align} 
where 
\begin{align}
 \la^Z_{aXY} := \frac{2c_W^2}{g^2} \left[g^Z_{d_X} \right]^*_{a2}
                                   \left[g^Z_{d_{Y}} \right]_{a3}, 
\quad 
 \la^H_{aXY} := \frac{v_H^2}{m_H^2} y_{Xa}^\dag y_{Y a}.  
\end{align}
The loop functions are defined as 
\begin{align}
 E_V(r) :=&\ 
\frac{1}{9} 
 -\frac{2}{3} \log r 
 - \frac{r(r^2+11r-18)}{12(1-r)^3}
 + \frac{r^2(4r^2-16r +15)}{6(1-r)^4} \log r, 
\\  
E_V^\pr(r) :=&\ 
- \frac{2}{3} 
 -\frac{r(r^2-5r-2)}{4(1-r)^3} + \frac{3}{2} \frac{r^2\log r}{(1-r)^4},
\\ 
E_V^\ppr(r) :=&\ 
     2 + \frac{3r(3-r)}{2(1-r)^2} + \frac{3r\log{r}}{(1-r)^3}, 
\end{align}
and 
\begin{align}
 E_S(r) := &\ -\frac{4}{9} 
  -\frac{1}{3}\log r 
  -\frac{r(19-41r+16r^2)}{36(1-r)^3}
  + \frac{r(-5+12r-8r^2+2r^3)}{6(1-r)^4} \log r, 
\\ 
E_S^\pr(r) :=&\ 
 \frac{1}{6} 
 + \frac{r(11-7r+2r^2)}{12(1-r)^3} + \frac{r\log r}{2(1-r)^4},  
\\    
E_S^\ppr(r) :=&\ -\frac{3}{2} 
  -\log r + \frac{r(-5+3r)}{2(1-r)^2} - \frac{r(3-3r+r^2)}{(1-r)^3} \log r. 
\end{align}
Here, $E_V$ and $E_V^\pr$ match to the Inami-Lim functions 
up to the constant terms.

The amplitudes for $b\to sg$ can be obtained by foramally replacing 
$eQ_d \to g_s T^\alpha$, 
where $T^\alpha$ is the generator of $SU(3)_C$ with $\alpha = 1,2,\cdots,8$.   
Note that the $W$ boson loop also has the same form, 
i.e. the formula can be applied for $B = W, Z, H$ for the gluon vertex.

\subsection{Wilson coefficients} 

The Wilson coefficients can be derived by matching with the amplitudes. 
$C_7$ and $C_8$ from the vector-like quarks are given by 
\begin{align}
 C_7^Q(\mu_Q) =&\ -\frac{1}{2} \sum_{a=4,5} \Biggl[
   D_W^\pr(r_{aW}) \left(\frac{\la^W_{aLL}}{\la^W_{3LL}}
                        +\frac{m_s}{m_b} \frac{\la^W_{aRR}}{\la^W_{3LL}} \right)
  +D_W^{\pr\pr}(r_{aW})   \frac{m_{U_a}}{m_b} \frac{\la^W_{aLR}}{\la^W_{3LL}} 
\\ \notag 
&\ + Q_d \sum_{B=Z,h} \left\{
   E_B^\pr(r_{aZ}) \left( \frac{\la^B_{aLL}}{\la^W_{3LL}}
                         +\frac{m_s}{m_b}  \frac{\la^B_{aRR}}{\la^W_{3LL}} \right)
  + E_B^{\pr\pr}(r_{aZ})\frac{m_{D_a}}{m_b}\frac{\la^B_{aLR}}{\la^W_{3LL}}  
\right\}
\Biggr], 
\\ 
 C_8^Q(\mu_Q) =&\ -\frac{1}{2} \sum_{a=4,5} \sum_{B=W,Z,h} \left\{
    E_B^\pr(r_{aZ}) \left(\frac{\la^B_{aLL}}{\la^W_{3LL}}
                        +\frac{m_s}{m_b} \frac{\la^B_{aRR}}{\la^W_{3LL}} \right)
  + E_B^{\pr\pr}(r_{aZ})\frac{m_{Q_a}^B}{m_b} \frac{\la^B_{aLR}}{\la^W_{3LL}} 
\right\},
\label{eq-C8VLQ}
\end{align}
where $E^{\pr,\ppr}_{Z,W} := E_V^{\pr,\ppr}$ and 
$m^B_{Q_a} := m_{U_a}~(m_{D_a})$ for $B=W$ ($Z,h$).  
Those for the primed ones are obtained by replacing $\LcR$ 
except $\la^W_{3LL}$ in the denominators.  
Here, $\mu_Q$ is a scale of the vector-like quarks 
which is assumed to be at TeV scale. 
The Wilson coefficients from the SM quark loops, denoted as $C_{7,8}^q$, 
are the same form, but the summation is changed to $a=1,2,3$ 
and these are matched at the EW scale $\mu_W := m_W$. 
We define the SM quark loop contribution after subtracting the SM one as 
\begin{align}
 \Delta C_7^q(\mu_W) := C_7^q +\frac{1}{2}\left(D_W^\pr(r_{aW})+\frac{23}{18}\right),
\quad
 \Delta C_8^q(\mu_W) := C_8^q +\frac{1}{2}\left(E_W^\pr(r_{aW}) +\frac{2}{3} \right), 
\end{align} 
where we subtract the constant terms in the loop functions 
which are canceled by the GIM mechanism in the SM.

We evaluated the other four-Fermi operators at the tree-level. 
The Wilson coefficients are calculated as 
\begin{align}
 C_2(\mu_W) = -\frac{\la^W_{2LL}}{\la^W_{3LL}}, 
\quad 
 C_{10}(\mu_W) = -\frac{m_c}{m_b} \frac{\la^W_{2RL}}{\la^W_{3LL}},  
\end{align}
and 
\begin{align}
 C_2^q(X,Y;\mu_W) = -\frac{2c_W^2}{g^2\la^W_{3LL}} 
                     \left[g^Z_{X}\right]_{sb}\left[g^Z_Y\right]_{qq}, 
\quad 
\hat{C}_2^d(X,Y;\mu_W) = - \frac{2c_W^2}{g^2\la^W_{3LL}} 
                     \left[g^Z_{X}\right]_{sd}\left[g^Z_Y\right]_{db}. 
\end{align}
We also define $\Delta C_2 := C_2-C_2^{\SM}$, 
with $C_2^\SM=0.991+0.0182i$,  
by subtracting the SM contribution as for $C_{7,8}$~\footnote{
With the Wolfenstein parameters, 
\begin{align}
 C_2^{\SM} = -\frac{V_{cs}^*V_{cb}}{V_{ts}^*V_{tb}} 
           = 1 - (\ol{\rho}-i\ol{\eta}) \la^2 
         -\left\{\frac{3}{8} -\left(\frac{1}{2}-\ol{\rho}+i\ol{\eta}\right)^2\right\}
          \la^4 
         + \order{\la^6}.   
\end{align}
}.

We are interested in the value of $C_7$ at $\mu_b = 2.5~\GeV$
in order to calculate $b\to s\gamma$. 
The new physics contributions to $C_7$ and $C_7^\pr$ are calculated as~\cite{Buras:2011zb} 
\begin{align}
\label{eq-DeltaC7}
\Delta C_7(\mu_b) =&\ 
       \sum_{i=7,8} 
       \left(\kappa_i^Q C^Q_i(\mu_Q) + \kappa_i^q \Delta C^q_i(\mu_W) \right) 
  \\ \notag 
    &\  \hspace{-1.5cm}
        +  \kappa_2 \Delta C_2(\mu_W) + \kappa_{10} C_{10}(\mu_W) 
        + \sum_{a=1,2} \sum_{X=L,R}\left(
                      \hat{\kappa}_{LX}^{a,d} \hat{C}_a^d(L,X;\mu_Z) 
                  +\sum_{q} \kappa_{LX}^{a,f} C_a^q(L,X;\mu_Z)  \right),
\\ 
C_7^\pr (\mu_b) =&\ 
       \sum_{i=7,8} 
     \left(\kappa_i^Q C^{Q\pr}_i(\mu_Q) + \kappa_i^q C^{q\pr}_i(\mu_W) \right) 
  \\ \notag 
    &\ \hspace{-1.5cm}
       +  \kappa_2 C^\pr_2(\mu_W) + \kappa_{10} C^\pr_{10}(\mu_W) 
       + \sum_{a=1,2}\sum_{X=L,R}\left(
                     \hat{\kappa}_{RX}^{a,d} \hat{C}_a^{d}(R,X;\mu_Z) 
                   + \sum_{q} \kappa_{RX}^{a,f} C_a^{q}(R,X;\mu_Z)  \right),
\end{align}
where the coefficients 
$\kappa_i^{q,Q}$, $\kappa_{2,10}$, $\hat{\kappa}^d_{LA}$ 
and $\kappa^q_{LA}$ are the RG running effects. 
The values of these coefficients and their derivations are summarized 
in App.~\ref{app-BsGamma}.
We can also calculate $C_8$ and $C_8^\pr$ in the same manner 
and we denote the RG coefficients for these by $\rho$'s instead of $\kappa$'s.

\subsection{Approximate formula} 

We discuss the leading contributions to flavor violations 
from the fourth-family vector-like quarks 
for the small Yukawa coupling constants $\eps^{q}_{X_i}$ 
and the heavy vector-like quarks, $m_{Q,U,D} \gg v_H$.  
We here summarize the results which are important in the following discussions. 
The details are shown in App.~\ref{app-analytical}.

We separate the flavor violating part of the $Z$ and Higgs boson couplings as 
\begin{align}
\label{eq-Zdevdef}
 \left[g^Z_{q_X}\right]_{ij} =: \frac{g}{c_W}\left[ 
\left(I_q \delta_{XL} - s_W^2 Q_q\right) \delta_{ij} + I_q \Delta^{q_X}_{ij} 
 \right], 
\quad 
 \left[Y^h_q \right]_{ij} = 
   \frac{1}{\sqrt{2}} \left(y^q_i \delta_{ij} + \delta y^q_{ij}\right),
\end{align} 
for $i,j=1,2,3$, where $y_i^q$ is the diagonal matrix 
whose elements are singular values of $y^{q}_{ij}$ 
appeared in the original Lagrangian Eq.~\eqref{eq-LNP}.  
Similarly, we also define the deviation of the $W$ boson couplings from the SM 
by the mixing with the fourth family as
\begin{align}
\left[g^W_{L}\right]_{ij} = \frac{g}{\sqrt{2}} \left(u_{ij} + \delta_{L} u_{ij}\right),
\quad 
\left[g^W_{R}\right]_{ij} = \frac{g}{\sqrt{2}} \delta_{R} u_{ij},
\end{align}
where $u_{ij}$ is the $3\times 3$ matrix that diagonalizes $y^d_{ij}$.  
Using the results of App.~\ref{app-analytical}, 
the deviations are approximately given by 
\begin{align}
 \Delta^{q_L}_{ij} \simeq -\teps^{q*}_{L_i} \teps^q_{L_j}  \frac{v_H^2}{m^2_{R_q}},
\quad   
 \Delta^{q_R}_{ij} \simeq  \teps^{q}_{R_i} \teps^{q*}_{R_j} \frac{v_H^2}{m^2_{Q}}, 
\end{align}
\begin{align}
\label{eq-YdevApp}
\delta y^q_{ij}    
\simeq&\
   2\ol{\la}_q \teps_{R_i}^q \teps^q_{L_j} \frac{v_H^2}{m_{Q}m_{R_q}},  
\end{align}
with $m_{R_u} := m_U$ and $m_{m_{R_d}} := m_D$. 
Here, 
\begin{align}
\teps^u_{L_i} := \eps^u_{L_i}, 
\quad 
\teps^u_{R_i} := \eps^u_{R_i}, 
\quad 
\teps^d_{L_i} := \eps^d_{L_k} u_L^{ki}, 
\quad  
\teps^d_{R_i} := (u_R^\dag)_{ik} \eps^d_{R_k}.  
\end{align}
The $W$ boson couplings are 
\begin{align}
\label{eq-WdevApp}
 \delta_L u_{ij} \simeq -\frac{1}{2}  
 \left(  \eps^{u*}_{L_i} \heps^u_{L_j}  \frac{v_H^2}{m_{U}^2}  
       + \eps^{d*}_{L_i} \teps^{d}_{L_j} \frac{v_H^2}{m_{D}^2}\right),
\quad 
 \delta_R u_{ij} \simeq \eps^u_{R_i} \teps^{d*}_{R_j} \frac{v_H^2}{m_{Q}^2},   
\end{align}
where $\heps^u_{L_{i}} := \eps^u_{L_k} u^L_{ki}$. 
The tree-level flavor violations by the $Z$ and Higgs bosons 
are induced at $\order{\eps^2 v_H^2/m_Q^2}$. 
In general, $\delta_L u$ induces non-unitarity of the CKM matrix 
for the three generations accompanied with the flavor violations of the $Z$ boson.

In App.~\ref{app-analytical}, 
we derived the approximate formula of the Wilson coefficients as  
\begin{align}
\label{eq-C7app} 
C_7^Q(\mu_Q) \simeq&\ \frac{v_H^2}{2m_Qm_D} 
\left\{\frac{11}{18}\frac{m_Qm_D}{m_U^2} \frac{\heps^{u*}_{L_2}\heps^u_{L_3}}{\la^W_{3LL}}
 +\frac{Q_d}{6} \frac{m_Q}{m_D} \frac{\teps^{d^*}_{L_2}\teps^d_{L_3}}{\la^W_{3LL}} 
 + \frac{v_H}{m_b}\ola_d^*\left(\frac{5}{6}+Q_d\right) 
   \frac{\teps^{d*}_{L_2}\teps^{d*}_{R_3}}{\la^W_{3LL}}
 \right\},
\\  
C_7^{Q\pr}(\mu_Q) \simeq &\ \frac{v_H^2}{2m_Qm_D}
 \left\{ 
  \left(\frac{11}{18}+\frac{Q_d}{6}\right) 
\frac{m_D}{m_Q} \frac{\teps^{d}_{R_2}\teps^{d*}_{R_3}}{\la^W_{3LL}}
+ \frac{v_H}{m_b} \ol{\la}_d \left( \frac{5}{6}+Q_d \right)
  \frac{\teps^d_{R_2}\teps^{d}_{L_3}}{\la^W_{3LL}}
 \right\}. 
\end{align}
The last terms in the parentheses are the chiral enhanced contributions. 
We estimate the deviation by the vector-like quarks as 
\begin{align}
 \Delta C_7^Q(\mu_b)  
                   \sim  0.011 \times 
                   \left(\frac{\ka^Q_7}{0.5}\right)
                   \left(\frac{0.0415}{\la^W_{3LL}}\right)
                   \left(\frac{\heps^{u*}_{L_2}\heps^u_{L_3}}{0.1}\right) 
                   \left(\frac{1~\TeV}{m_U}\right)^2,  
\end{align}
from the first term in Eq.~\eqref{eq-C7app}, and  
\begin{align}
 \Delta C_7^Q(\mu_b)  
                   \sim  0.11 \times  \ola_d^* 
                   \left(\frac{\ka^Q_7}{0.5}\right)
                   \left(\frac{0.0415}{\la^W_{3LL}}\right)
                   \left(\frac{\teps^{d*}_{L_2}\teps^d_{R_3}}{0.01}\right) 
                   \left(\frac{1~\TeV}{\sqrt{m_Qm_D}}\right)^2,  
\end{align}
from the chiral enhanced ones. 
Thus the latter can be $\order{10\%}$ of the SM contribution 
if these sizes of the mixing angles are consistent with the other constraints. 
In the next section, 
we shall study several constraints to the tree-level flavor violations, 
and examine how the chiral enhanced effects can be sizable.

\section{Phenomenology} 
\label{sec-pheno}

We evaluate the effect of the vector-like fourth family 
to the flavor violating observables, especially $b\to s\gamma$.  
Throughout this work, we take 
\begin{align}
 \eps^u_{L_1} =  \eps^u_{R_1} = \eps^d_{L_1} =  \eps^d_{R_1} = 0,   
\end{align}
so that there is no flavor violation involving the down quark.  
In addition to $b\to s\gamma$, 
we examine the observables in top quark decays, 
$Z$-pole observables and $B_s$-$\ol{B}_s$ mixing.

\subsection{\texorpdfstring{$\overline{B}\to X_s\gamma$}{Bbar -> Xs gamma}}

We evaluate the branching fraction $\ol{B}\to X_s \gamma$ 
from the Wilson coefficients $C_7$ and $C_7^\prime$ derived in the previous section. 
We separate the SM and new physics contributions as 
\begin{align}
 \br{\ol{B}}{X_s \gamma}
=&\  \br{\ol{B}}{X_s \gamma}_{\SM} + R \left(
 2C_7^{\SM}\,\mathrm{Re}\,\left(\Delta C_7 \right) +\abs{\Delta C_7}^2 
      + \abs{\Delta C_7^\prime}^2 
 \right), 
\end{align}
where the Wilson coefficients are evaluated at $\mu_b = 2.5~\GeV$.  
The SM prediction with NNLO QCD accuracy is~\cite{Misiak:2015xwa,Misiak:2020vlo}  
\begin{align}
 \br{\ol{B}}{X_s\gamma}_{\SM} = \left(3.40\pm 0.17\right)\times 10^{-4}. 
\end{align}
The SM Wilson coefficient $C_7^{\SM}$ and the numerical factor $R$
are given by~\cite{Gambino:2001ew,Misiak:2006ab,Misiak:2006zs} 
\begin{align}
R=2.47\times 10^{-3}, 
\quad  
C_7^{\SM}(\mu_b) =  -0.3523. 
\end{align}  
The new physics contributions to $\Delta C_7$ and $C_7^\pr$ 
are defined in Eq.~\eqref{eq-DeltaC7}. 
The measured value is~\cite{ParticleDataGroup:2024cfk,HeavyFlavorAveragingGroupHFLAV:2024ctg}
\begin{align}
 \br{\ol{B}}{X_s\gamma} = \left(3.49\pm 0.19\right)\times 10^{-4}. 
\end{align}

The CP violation can be induced in our setup 
by the complex phases of the Yukawa couplings.  
Belle and BaBar measure the CP asymmetry~\cite{Belle:2018iff}
\begin{align}
 A_{\CP} := 
\frac{\Gamma\left(\ol{B}\to\ol{X}_s\gamma\right)-
      \Gamma\left(B\to X_s\gamma\right)} 
     {\Gamma\left(\ol{B}\to\ol{X}_s\gamma\right)+
      \Gamma\left(B\to X_s\gamma\right)}.  
\end{align}
This quantity has, however, 
significant theoretical uncertainty from the resolved photon contribution.  
To avoid it, we can consider the new quantity defined as~\cite{Benzke:2010tq} 
\begin{align}
\label{eq-dACP}
 \Delta A_{\CP} := 
A_\CP\left(B^+ \to X_s^+ \gamma\right)-A_\CP\left(B^0\to X_s^0 \gamma\right).  
\end{align}
This can be evaluated semi-numerically as 
\begin{align}
\label{eq-ACP}
 \Delta A_{\CP} 
             \sim 0.12\times \left(\frac{\tilde{\Lambda}_{78}}{100~\MeV}\right)
             \left[\mathrm{Im}\left(\frac{C_8}{C_7}\right)
             + 
            \abs{\frac{C_7^\prime}{C_7}}^2 
            \mathrm{Im}\left(\frac{C_8^\prime}{C_7^\prime}\right) 
           \right], 
\end{align} 
where $\tilde{\Lambda}_{78}$ is the hadronic parameter 
denoting the interference between electomagnetic 
and chromomagnetic dipole diagrams whose range is estimated as 
$\tilde{\Lambda}_{78} \in [17, 190]~\MeV$~\cite{Lee:2006wn,Benzke:2010js}. 
The BaBar~\cite{BaBar:2014czi} and Belle~\cite{Belle:2018iff} 
results are respectively given by 
\begin{align}
 \Delta A_{\CP} = (+5.0\pm 3.9 \pm 1.5) \%, 
\quad 
 \Delta A_{\CP} = (+3.69\pm 2.65 \pm 0.76) \%, 
\end{align}
which are both $\order{5\%}$ accuracy.

\subsection{CKM elements and charged right-current interaction}

With the fourth family, the CKM matrix is not unitary, 
so we refer the values of the individual elements~\cite{ParticleDataGroup:2024cfk}
without assuming the unitarity,  
\begin{align}
\label{eq-absCKMexp}
 \abs{V_{ud}} =&\ 0.97367 \pm 0.00032, 
\quad
& \abs{V_{us}} =&\ 0.22431 \pm 0.00085, 
\quad 
& \abs{V_{ub}} =&\ (3.82 \pm 0.20)\times 10^{-3}, 
\notag \\ \notag 
\abs{V_{cd}} =&\ 0.221 \pm 0.004, 
\quad 
& \abs{V_{cs}} =&\ 0.975 \pm 0.006, 
\quad 
& \abs{V_{cb}} =&\ (41.1 \pm 1.2)\times 10^{-3}, 
\\
\abs{V_{td}} =&\ (8.6\pm 0.2)\times 10^{-3}, 
\quad 
& \abs{V_{ts}} =&\ (41.5\pm 0.9)\times 10^{-3}, 
\quad 
& \abs{V_{tb}} =&\ 1.010\pm 0.027, 
\end{align}
and the CP angles 
\begin{align}
 \sin 2\beta = 0.709 \pm 0.011, 
\quad 
 \alpha = (84.1^{+4.5}_{-3.8})^{\circ}, 
\quad 
 \gamma = (65.7\pm3.0)^{\circ}.  
\end{align}
Here, the angles are defined as 
\begin{align}
\label{eq-phaseCKMexp}
 \alpha = \mathrm{Arg}\left( 
-\frac{V_{cd}V_{cb}^*}{V_{td}V_{tb}^*}
\right), 
\quad 
 \beta = \mathrm{Arg}\left( 
-\frac{V_{td}V_{tb}^*}{V_{ud}V_{ub}^*}
\right), 
\quad 
 \gamma = \mathrm{Arg}\left( 
-\frac{V_{ud}V_{ub}^*}{V_{cd}V_{cb}^*}
\right). 
\end{align}
Note that $\alpha + \beta + \gamma = 180^{\circ}$ 
is ensured by the unitarity in the SM, while it is not in our case.   
In this work, we do not consider how the vector-like quarks 
affect to the CKM fit, and leave it as our future work.   
Whereas, we restrict the parameter space such that these values are reproduced within the experimental uncertainties given in Eqs.~\eqref{eq-absCKMexp} and~\eqref{eq-phaseCKMexp}, by fitting the parameters in $y_i^{u}$ and $y^d_{ij}$ of Eq.~\eqref{eq-LNP}.

There are also the right-handed interactions of the $W$ boson,
which would affect to the $W$ boson polarization of the top quark decay. 
We define the fractional quantities~\cite{ParticleDataGroup:2024cfk},  
\begin{align}
 F_0 := \frac{\Gamma_{0}}{\Gamma_{t}}, 
\quad 
 F_L := \frac{\Gamma_{L}}{\Gamma_{t}}, 
\quad 
 F_R := \frac{\Gamma_{R}}{\Gamma_{t}}, 
\end{align}
where $\Gamma_{t}$ is the total width of the top quark. 
The indices represent the polarization of the $W$ boson. 
These are evaluated as  
\begin{align}
 F_0 = \frac{m_t^2}{m_t^2+2m_W^2}, \quad 
 F_X = \frac{\abs{\left[g^W_{X}\right]_{33}}^2 2m_W^2}
           {\left(
             \abs{\left[g^W_{L}\right]_{33}}^2 
           + \abs{\left[g^W_{R}\right]_{33}}^2 \right) (m_W^2+2m_{t}^2)},   
\end{align}
where $X=L,R$.  
The latest ATLAS result is~\cite{Aguilar-Saavedra:2010ljg,Aguilar-Saavedra:2015yza},  
\begin{align}
&\ F_0 = 0.684  \pm 0.005 \pm 0.014, \\ \notag   
&\ F_L = 0.318  \pm 0.003 \pm 0.008, \\ \notag   
&\ F_R = -0.002 \pm 0.002 \pm 0.014, 
\end{align}
where the first (second) uncertainty 
is from statistics (systematics)~\cite{ATLAS:2022rms}.

\subsection{\texorpdfstring{$Z$}{Z}-pole observables} 

The diagonal $Z$ boson couplings are also deviated from the SM predictions.  
We define the (axial) vector $Z$ boson couplings as 
\begin{align}
 g^{q_i}_V := \frac{1}{2}
           \left(\left[g^Z_{q_L}\right]_{ii}+\left[g^Z_{q_R}\right]_{ii}\right),
\quad 
 g^{q_i}_A := \frac{1}{2}
           \left(\left[g^Z_{q_L}\right]_{ii}-\left[g^Z_{q_R}\right]_{ii}\right),  
\end{align}
where $(q,i) = (u,2)$ for the charm quark $q_i = u_2 = c$. 
The bottom quark is labeled in the same manner as $d_3 = b$.  
The asymmetry parameters are calculated as
\begin{align}
 A_{q_i} :=&\ 
   \frac{2 g^{q_i}_V g^{q_i}_A}{{(g^{q_i}_V})^2 + ({g^{q_i}_A})^2} 
\\ \notag   
  \simeq&\ \frac{1-4s_W^2\abs{Q_q}}{1-4s_W^2\abs{Q_q}+8s_W^4\abs{Q_q}^2}  
\\ \notag 
&\ + \frac{8s_W^2\abs{Q_q}\left(1-2s_W^2\abs{Q_q}\right)}
       {\left(1-4s_W^2\abs{Q_q}+8s_W^4\abs{Q_q}^2\right)^2 }
  \left\{ 2s_W^2 \abs{Q_q} \Delta^{q_L}_{ii} 
        +(1-2s_W^2\abs{Q_q}) \Delta^{q_R}_{ii} 
      \right\}. 
\end{align}  
where $\Delta^{q_A}_{ii}$ is the deviation of the $Z$-boson coupling 
defined in Eq.~\eqref{eq-Zdevdef},  
Hence, 
\begin{align}
 A_c \sim&\ 0.673 + 0.789 \times \Delta^{u_L}_{22} + 1.78 \times \Delta^{u_R}_{22},
\\ \notag 
 A_b \sim&\ 0.937 + 0.145 \times \Delta^{d_L}_{33} + 0.802 \times \Delta^{d_R}_{33},  
\end{align}
where $s_W^2 = 0.2299 + 0.00032$ is used. 
The measured values are~\cite{ParticleDataGroup:2024cfk}  
\begin{align}
A_c = 0.670 \pm 0.027, \quad  A_b = 0.923\pm 0.020. 
\end{align}
In general, the deviation 
\begin{align}
 \Delta^{q_A}_{ii} \sim \abs{\eps^{q}_{A_i}}^2 \frac{v_H^2}{m_Q^2} 
                   \sim 0.03\times \abs{\eps^{q}_{A_i}}^2
                                   \left(\frac{1~\TeV}{m_Q}\right)^2,  
\end{align}
is sufficiently small for $\eps^{q}_{A_i} \lesssim \order{1}$.

\subsection{Flavor violating top decays} 

The top quark can decay to $Z/H$ and the charm quark at the tree-level.  
Neglecting the charm quark mass, the decay widths are given by 
\begin{align}
 \Gamma(t\to cZ) =&\ \frac{m_t^3}{32\pi m_Z^2} 
                     \left(\abs{\left[g^Z_{u_L}\right]_{23}  }^2 
                         + \abs{\left[g^Z_{u_R}\right]_{23}}^2 \right)
                     \left(1+2\frac{m_Z^2}{m_t^2}\right)
                     \left(1- \frac{m_Z^2}{m_t^2}\right)^2, 
\\ \notag 
\Gamma(t\to ch) =&\ \frac{m_t}{32\pi} 
                     \left(\abs{\left[Y^h_{u}\right]_{23}}^2 
                         + \abs{\left[Y^h_{u}\right]_{32}}^2 \right)
                     \left(1- \frac{m_H^2}{m_t^2}\right)^2.   
\end{align}
The branching fractions are estimated as 
\begin{align}
 \br{t}{cZ}  \sim&\ \frac{G_Fm_t^3}{8\sqrt{2}\pi\Gamma_t} 
                     \left(1+2\frac{m_Z^2}{m_t^2}\right)
                     \left(1- \frac{m_Z^2}{m_t^2}\right)^2 
                     \max_{X=L,R}\left( \abs{\Delta^{u_X}_{23}}^2 \right)   
 \\ \notag 
              \sim&\ 9.4\times 10^{-5} \times 
                            \max_{X=L,R}
                 \left(\frac{\abs{\eps^{u}_{X_2}\eps^{u}_{X_3}}^2}{0.1}\right)   
                 \left(\frac{1~\TeV}{m_\VLQ}\right)^4, 
\\ \notag 
 \br{t}{ch} \sim&\  \frac{G_Fm_t^3}{64\sqrt{2}\pi\Gamma_t} \frac{4v_H^2}{m_t^2} 
                    \left(1-\frac{m_H^2}{m_t^2}\right)
                  \max \left( \abs{\delta y^u_{23}}^2, \abs{\delta y^u_{32}}^2 \right)
  \\ \notag 
            \sim&\ 2.7\times 10^{-4} \times 
                            \max_{X=L,R}
                 \left(\frac{\abs{\eps^{u}_{X_2}\eps^{u}_{X_3}}^2}{0.1}, 
                       \frac{2v_H}{m_t}
                       \frac{\abs{\eps^{u}_{X_2}\eps^{u}_{\ol{X}_3}}^2}{0.1}
                      \right)   
                 \left(\frac{1~\TeV}{m_\VLQ}\right)^4, 
\end{align}
where $\ol{X} = R,L$ for $X=L,R$, respectively. 
$\Gamma_t = 1.326~\GeV$ is the total width of the top quark~\cite{ParticleDataGroup:2024cfk}. 
The experimental bounds are~\cite{ParticleDataGroup:2024cfk}
\begin{align}
 \br{t}{cZ} < 1.2\times 10^{-4}, 
\quad 
 \br{t}{ch} < 3.4\times 10^{-4}.  
\end{align}
Thus, $\eps^u_{L,R} < \order{1}$ 
is sufficiently small for the vector-like quarks at TeV scale.

Similarly to  $b\to s\gamma$, 
the vector-like quarks also induce $t \to c\gamma$ decay  
through the dipole operator,  
\begin{align}
\Lcal_{t,\mathrm{dipole}}  
   = \frac{ie}{4\pi^2} m_t 
                \ol{c} \sigma^{\mu\nu}\left(
                C_L^{ut} P_R + C_R^{ut} P_L   \right)t F_{\mu\nu},  
\end{align}
analogous to $Q_7$ and $Q_7^\pr$, up to their normalizations.  
The decay width is given by 
\begin{align}
 \Gamma\left(t\to c\gamma \right)
= \frac{\alpha_e m_t^5}{16\pi^4} \left( \abs{C_L^{ut}}^2+\abs{C_R^{ut}}^2 \right). 
\end{align}
The 95\% C.L. bound is~\cite{ParticleDataGroup:2024cfk} 
\begin{align}
  \br{t}{c\gamma} < 9.5\times 10^{-6}.  
\end{align}

The Wilson coefficients are given by 
\begin{align}
 C_L^{ut} =&\ \frac{G_F}{\sqrt{2}} \sum_{a=1}^5 \Biggl[ 
        \tD_W^\pr(r_{aW}) \tla^W_{aLL} 
       +\frac{m_{D_a}}{m_t}\tD_W^{\pr\pr}(r_{aW}) \tla^W_{aLR} 
\\ \notag 
   & \hspace{2.0cm} 
  + Q_u \sum_{B=Z,h} \left(  
   E^\pr_{B}(r_{aB}) \tla^B_{aLL}+\frac{m_{U_a}}{m_t}E^{\pr\pr}_B(r_{aB}) \tla^B_{aLR} 
    \right) 
    \Biggr],  
\end{align}
where 
\begin{align}
 \tla^W_{aXY} := 
      \frac{2}{g^2} \left[g^W_{X}\right]_{ca} \left[g^W_{Y}\right]_{ta}^*, 
\quad 
 \tla^Z_{aXY} := 
 \frac{2c_W^2}{g^2} \left[g^Z_{u_X}\right]_{ca} \left[g^Z_{u_{Y}}\right]_{ta}^*, 
\quad 
 \tla^h_{aXY} := \frac{v_H^2}{m_H^2} y_{Xa}^\dag y_{Y a}.  
\end{align}
Here $y^\dag_{Aa}$ and $y_{Aa}$ are defined in the same way as Eq.~\eqref{eq-yLRa}, 
but for $Y^h_u$ instead of $Y^h_d$.  
The loop functions for the $W$ boson loop are changed to 
\begin{align}
  \tD^\pr_W(x) :=&\ \frac{19}{18} 
                    -\frac{x(-5+10x+7x^2)}{12(x-1)^3}
                    +\frac{x^2(-1+3x)\log x}{2(x-1)^4}, 
\\ 
  \tD^{\pr\pr}_W(x) :=&\ 
                    -\frac{8}{3} 
                    +\frac{2x^2}{(x-1)^2}  
                    +\frac{x(1-3x)\log x}{(x-1)^3},  
\end{align}
due to the difference of the electric charge. 
The expression for $C^{ut}_{R}$ is obtained by formally replacing $\LcR$. 
Note that the chiral enhancement factor $v_H/m_t$ 
is not large unlike the case of $b\to s\gamma$. 
Assuming that the chiral enhanced effects of the $W$ boson loop dominates, 
\begin{align}
 \br{t}{c\gamma} \sim&\ \frac{\alpha_e m_t^3 G_F}{64\sqrt{2}\pi^4\Gamma_t}
                        \abs{\ol{\la}_d \tD^{\pr\pr}_W(r_{aW}) }^2 
                        \abs{\tla^W_{aLR}}^2 
 \\ \notag 
               \sim&\ 4.0\times 10^{-5} \times   
                        \abs{\ol{\la}_d \tD^{\pr\pr}_W(r_{aW})}^2 
                        \abs{\tla^W_{aLR}}^2.  
\end{align}
This is smaller than the experimental bound of $\order{10^{-5}}$
even without the suppression by the small mixing angles in $\tla^W_{aLR}$.   
Thus $t\to c\gamma$ is always smaller than the experimental bounds 
for the heavy vector-like quarks.

\subsection{\texorpdfstring{$B_s$-$\ol{B}_s$}{Bs-Bsbar} mixing}

There are tree-level flavor violations through the $Z$ and Higgs bosons~\footnote{
The 1-loop calculation with the box contribution of the vector-like quark loops 
is our future work. 
}.  
In this work, we examine the limits from the $B_s$-$\ol{B}_s$ mixing.  
We shall work on the operator basis~\cite{Buras:2012fs},  
\begin{align}
\label{eq-OBBbar}
Q_1^{\VLL} =& \left(\ol{b} \gamma^\mu P_L s\right) 
                 \left(\ol{b} \gamma_\mu P_L s\right),  
\quad & 
Q_1^{\VRR} =& \left(\ol{b} \gamma^\mu P_R s\right) 
                 \left(\ol{b} \gamma_\mu P_R s\right),  
\\ \notag
Q_1^{\LR} =& \left(\ol{b} \gamma^\mu P_L s\right) 
                \left(\ol{b} \gamma_\mu P_R s\right),  
\quad & 
Q_2^{\LR} =& \left(\ol{b} P_L s\right) 
                \left(\ol{b} P_R s\right),  
\\ \notag 
Q_1^{\SLL} =& \left(\ol{b} P_L s\right) 
                 \left(\ol{b} P_L s\right),  
\quad & 
Q_1^{\SRR} =& \left(\ol{b} P_R s\right) 
                 \left(\ol{b} P_R s\right),  
\\ \notag 
Q_2^{\SLL} =& \left(\ol{b} \sigma^{\mu\nu} P_L s\right) 
                 \left(\ol{b} \sigma_{\mu\nu} P_L s\right),  
\quad & 
Q_2^{\SRR} =& \left(\ol{b} \sigma^{\mu\nu} P_R s\right) 
                 \left(\ol{b} \sigma_{\mu\nu} P_R s\right),  
\end{align}
where the color indices are contracted inside each parenthesis.

The off-diagonal element of the neutral meson anti-meson system 
is given by 
\begin{align}
 M_{12}^*(B_s):= \bra{\ol{B}_s} \mathcal{H}_{\mathrm{eff}} \ket{B_s} 
          =  M_{12}^{\mathrm{SM}*} + \Delta M_{12}(B_s), 
\end{align}
where $M^\SM_{12}$ is the SM contribution.  
The new physics contribution $\Delta M_{12}$ is evaluated as 
\begin{align}
\label{eq-defPa}
2m_{B_s} \Delta M_{12}(B_s) = \sum_{a,\Gamma} C_a^\Gamma(\mu_L) \vev{Q_a}(\mu_L) 
         =: \sum_{a,\Gamma} P_{a}^{\Gamma} (\mu_Z) C_a^\Gamma(\mu_Z),   
\end{align} 
where $(a,\Gamma) = (1,\VLL), (1,\LR), (2,\LR),  (1,\SLL), (2,\SLL)$. 
Here, the numerical coefficient $P_a^\Gamma(\mu_Z)$ 
includes the bag parameters and the RG running 
between the lattice scale $\mu_L$ and the weak scale $\mu_Z = m_Z$. 
Their values are calculated in App.~\ref{app-NMM}.  
Those for $Q_1^{\VRR}$ and $Q_{1,2}^{\SRR}$ 
are the same as $Q_1^{\VLL}$ and $Q_{1,2}^{\SLL}$, respectively.

The Wilson coefficients induced by the tree-level exchange 
of the $Z$ and Higgs bosons at NLO QCD are given by~\cite{Buras:2012fs}, 
\begin{align}
 C_1^{\VLL}(\mu) 
   =&\ \frac{g_Lg_L}{2m_Z^2} \left[1 + \frac{\alpha_s}{4\pi} 
       \left(-2 \log\frac{m_Z^2}{\mu^2} + \frac{11}{3}  \right)\right], 
\\ \notag 
 C_1^{\LR}(\mu) 
   =&\ \frac{g_Lg_R}{m_Z^2} \left[1 + \frac{\alpha_s}{4\pi} 
       \left(-\log\frac{m_Z^2}{\mu^2} - \frac{1}{6}  \right)\right]
       - \frac{y_Ly_R}{m_H^2}\left(-\frac{3}{2}\right) \frac{\alpha_s}{4\pi} 
      , 
\\ \notag 
 C_2^{\LR}(\mu) 
   =&\ \frac{g_Lg_R}{m_Z^2}  \frac{\alpha_s}{4\pi} 
       \left(-6\log\frac{m_Z^2}{\mu^2} - 1  \right)
       -\frac{y_Ly_R}{m_H^2}\left(1-\frac{\alpha_s}{4\pi}\right), 
\\ \notag 
 C_1^{\SLL}(\mu) 
   =&\ -\frac{y_Ly_L}{2m_H^2} \left[1 + \frac{\alpha_s}{4\pi} 
        \left(-3\log\frac{m_H^2}{\mu^2} + \frac{9}{2}  \right)\right], 
\\ \notag 
 C_2^{\SLL}(\mu) 
   =&\ -\frac{y_Ly_L}{2m_H^2}  \frac{\alpha_s}{4\pi} 
       \left(-\frac{1}{12}\log\frac{m_H^2}{\mu^2} + \frac{1}{8}  \right), 
\end{align}
where 
\begin{align}
 g_L := \left[ g^{Z}_{d_L} \right]_{bs}, 
\quad  
 g_R := \left[ g^{Z}_{d_R} \right]_{bs}, 
\quad 
 y_L := \left[ Y^{h}_d \right]_{bs}, 
\quad 
 y_R := \left[ Y^{h}_d \right]_{sb}^*.
\end{align}
We take the renormalization scale $\mu = m_Z$.  
The mass difference is defined as
\begin{align}
 \Delta M_s := 2 \abs{M_{12}(B_s)},  
\end{align}
whose observed value~\cite{HeavyFlavorAveragingGroupHFLAV:2024ctg} 
and the SM predictions~\cite{Albrecht:2024oyn,Bigi:2001ys,Artuso:2015swg}
are respectively given by 
\begin{align}
 (\Delta M_s)_{\mathrm{exp}} = 17.766 \pm 0.006~\mathrm{ps}^{-1}, 
 \quad 
 (\Delta M_s)_{\SM} = 18.23\pm 0.63~\mathrm{ps}^{-1}.  
\end{align}

When we focus on the tree-level flavor violation by the Higgs boson,  
the new physics effect is estimated as 
\begin{align}
\label{eq-RBs}
R_{B_s} := \abs{\frac{\Delta M_{12}^{B_s}}{M_{12}^{\SM}(B_s)}}
\sim 
0.60 \times \left(\frac{\delta y^d_{32}}{10^{-3}} \right)^2
\sim  
0.0056 \times \ol{\la}_d \left(\frac{\teps^d_{R_3}\teps^d_{L_2}}{0.01}\right)^2
                        \left(\frac{2.5~\TeV}{\sqrt{m_Qm_D}} \right)^4, 
\end{align}
whereas the theoretical uncertainty is about $3.5\%$. 
Thus the constraints on the tree-level $B_s$-$\ol{B}_s$ mixing 
is expected to be weaker than that on $\ol{B}\to X_s \gamma$
if there is the chiral enhancement effect. 
Here, we define the absolute value of the ratio in Eq.~\eqref{eq-RBs}, 
$R_{B_s}$ to be compared with the constraint from  
$\ol{B}\to X_s \gamma$ in the next section.

\subsection{LHC limits on vector-like quarks}

The LHC limits on the direct vector-like quarks 
are summarized by ATLAS in Ref.~\cite{ATLAS:2024fdw} 
and by CMS in Ref.~\cite{CMS:2024bni}. 
From the pair-production results, 
the lower bounds are 1-1.6 TeV depending on the decay modes. 
In our case, the single productions are expected to be suppressed 
due to the small off-diagonal couplings between $Z/W$ boson and quarks.   
Thus the vector-like quarks heavier than 2 TeV 
are far from the current bounds at the LHC.

\section{Numerical analysis}  
\label{sec-analysis}

\begin{figure}[!t]
\centering
\begin{minipage}{0.48\textwidth}
  \centering
  \includegraphics[width=\linewidth]{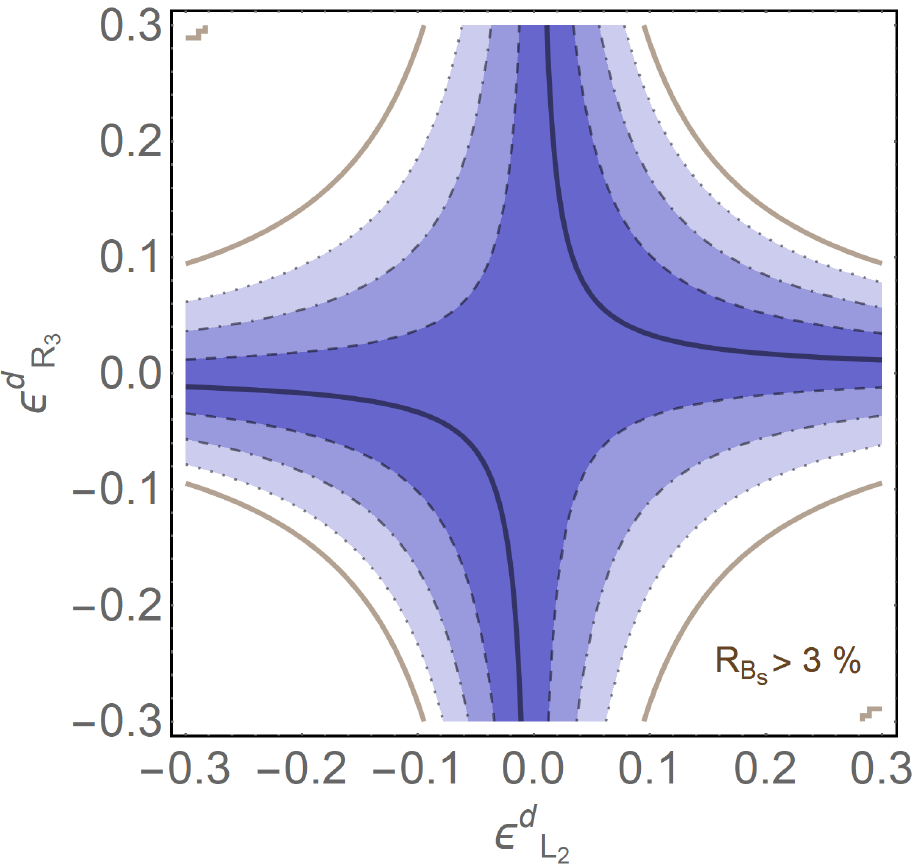}
\end{minipage}
\begin{minipage}{0.48\textwidth}
  \centering
  \includegraphics[width=\linewidth]{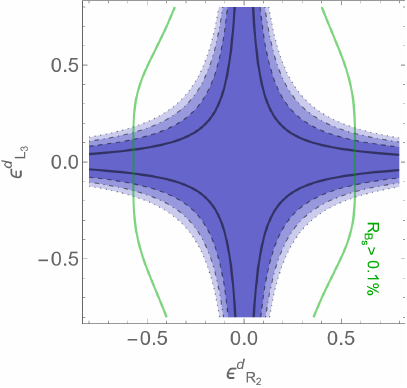}
\end{minipage}

\begin{minipage}{0.48\textwidth}
  \centering
  \includegraphics[width=\linewidth]{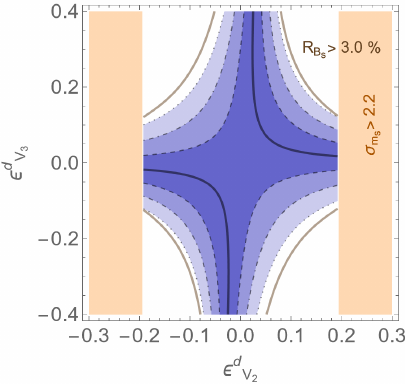}
\end{minipage}
\begin{minipage}{0.48\textwidth}
  \centering
  \includegraphics[width=\linewidth]{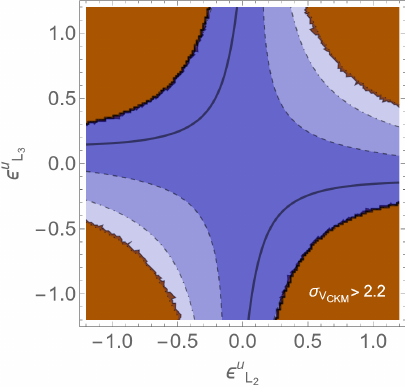}
\end{minipage}
\caption{\label{fig_bsg}  
 $\br{\ol{B}}{X_s\gamma}$ of the cases (1), (2), (3) and (4) 
 on the left-top, right-top, bottom-left and bottom-right panels, respectively.
 The thick black line is the experimental central value.
The deviations are $1\sigma$, $2\sigma$ and $3\sigma$ on the dashed, dot-dashed and dotted lines, respectively. $R_{B_s}$, that is the ratio of $\Delta M_{s}$ to the SM prediction, reaches 3 $\%$ (0.1 $\%$) on the brown (green) lines. 
 The quark mass and CKM fit is failed in the orange and brown colored region 
 in the bottom panels. 
 }
\end{figure}

\begin{figure}[!t]
\centering
\includegraphics[width=0.6\linewidth]{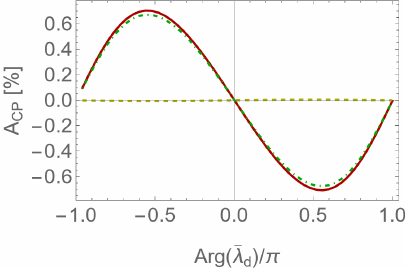}
\caption{\label{fig_ACP} 
$A_{\CP}$ with respect to $\mathrm{Arg}\ola_d$. 
The solid red, dot-dashed green and dashed yellow 
lines are respectively the cases  (1), (2) and (3) 
where $\eps^d_{A_{2,3}} = 0.1$ with $A=L,R,V$. 
}
\end{figure}

In our numerical analysis, 
we fix the vector-like mass parameters as 
\begin{align}
 m_Q = 3~\TeV, \quad m_U = m_D = 2~\TeV, 
\end{align}
and the Yukawa couplings as 
\begin{align}
 \la_u = \la_d = 0.1, \quad \ola_u = \abs{\ola_d} = 1.0. 
\end{align} 
The phase of $\ola_d$ will be varied later in the analysis of $A_{\CP}$.  
The SM Yukawa couplings are parametrized as 
\begin{align}
 y^u_{ij} = y^u_i \delta_{ij}, 
\quad  
 y^d_{ij} = \left(u_R \right)_{ik} y^d_k \left(u_L^\dag \right)_{kj}.  
\end{align}
The diagonal elements and $u_L$ are fit 
to the observed quark masses and the CKM angles. 
The target values of the quark masses evaluated at $\mu=m_Z$, calculated in Ref.~\cite{Antusch:2013jca}, 
and those of the CKM angles are shown in Eqs.~\eqref{eq-absCKMexp} and~\eqref{eq-phaseCKMexp}. 
For simplicity, we take $u_R$ is a $3\times 3$ identity matrix.  
We allow the deviations from the central values up to $2.2\sigma$,
because the CKM fit with the $3\times 3$ unitary matrix in the SM 
has the deviation of $\abs{V_{ud}}$ 
slightly higher than $2.0\sigma$~\cite{ParticleDataGroup:2024cfk}.

For the off-diagonal Yukawa couplings, we consider the following scenarios: 
\begin{enumerate}[label=(\arabic*)]
 \item{$(\eps^{d}_{L_2}, \eps^{d}_{R_3})$ are non-zero, and the others are zero.} 
 \item{$(\eps^{d}_{L_3}, \eps^{d}_{R_2})$ are non-zero, and the others are zero.} 
 \item{$(\eps^d_{V_2}:= \eps^{d}_{L_2} = \eps^{d}_{R_2}, 
         \eps^d_{V_3}:= \eps^{d}_{L_3} = \eps^{d}_{R_3})$ are non-zero, and the others are zero.} 
 \item{$(\eps^{u}_{L_2}, \eps^{u}_{L_3})$ are non-zero, and the others are zero.} 
\end{enumerate}
Although our main interest is (1)-(3), 
where the chiral enhancement occurs, 
we also study the flavor violations induced only from the up-quark sector, 
where the other constraints are very weak.

Figure~\ref{fig_bsg} shows the values of $\br{\ol{B}}{X_s\gamma}$ 
of the four scenarios: the predictions in the cases (1), (2), (3) and (4) are
 on the left-top, right-top, bottom-left and bottom-right panels, respectively.
Our prediction matches the experimental central value on the thick black line.
The deviations are $1\sigma$, $2\sigma$ and $3\sigma$ on the dashed, dot-dashed and dotted lines, respectively. 
The ratio $R_{B_s}$ reaches 3 $\%$ (0.1 $\%$) 
on the brown (green) lines in Fig.~\ref{fig_bsg}.

In the top panels, 
we could find a good fit to the quark mass and CKM elements
over the parameter space. 
On the top-left panel showing the case (1), 
the vector-like quark contributes only to $C_7$, see Eq.~\eqref{eq-C7app}.
The sign of $\Delta C_7$ affects the interference with the SM contribution. 
We see that the deviation of the $\br{\ol{B}}{X_s\gamma}$ is significant, 
while that of the $B_s$-$\ol{B}_s$ mixing at the tree-level is less than $3\%$. 
Thus the chiral enhancement effect 
can drastically affect to $\br{\ol{B}}{X_s\gamma}$ 
without changing the other observables. 
The top-right panel shows the case (2), where $\Delta C_7^\pr$ is induced.  
Since there is no interference with the SM one, 
the branching fraction is almost independent of the sign of the couplings. 
The deviation of the $B_s$-$\ol{B}_s$ mixing is negligible, 
because the flavor violating Yukawa coupling $\delta y^d_{32}$ 
is not induced in this case as we can read from the approximate formula Eq.~\eqref{eq-YdevApp}.

The case (3) is shown in the bottom-left panel. 
As shown in Eq.~\eqref{eq-WdevApp}, 
the CKM elements are deviated 
by the mixing effect proportional to $\eps^d_{L_a}\eps^d_{L_b}$ with $a,b = 2,3$,
as well as the quark masses through the Yukawa coupling in Eq.~\eqref{eq-YdevApp}.  
We could not find a good fit to the strange quark mass 
at $\abs{\eps^d_{V_2}} \gtrsim 0.25$, 
where the deviation of the Yukawa couplings are comparable to the SM one, 
\begin{align}
\frac{\delta y^d_{22}}{y_s} 
   \sim \ol{\la}_d 
        \left(\frac{\eps^d_{V_2}}{0.24}\right)^2 
        \left(\frac{3~\TeV}{m_Q}\right)
        \left(\frac{2~\TeV}{m_D}\right). 
\end{align}
Hence, the $\br{\ol{B}}{X_s\gamma}$ is the most sensitive 
at $\abs{\eps^d_{V_3}} \gtrsim \eps^{d}_{V_2}$, 
whereas the quark mass and CKM fit can be the most sensitive in the opposite regime.

The bottom-right panel shows the case (4), 
where there are non-zero mixing only in the up-type quarks. 
This case does not induce the chiral enhancement to $\br{\ol{B}}{X_s\gamma}$, 
so the vector-like quark contribution is typically smaller than the other cases. 
Nonetheless, we examine this case because the larger mixing couplings 
$\eps^u_{L_{2,3}}$ are allowed due to the absence of the relatively tight constraints in the down-type quarks.    
As for the case (3), the mixing also affects to the CKM elements, 
especially, 
the large couplings affect to the unitarity of the $3\times 3$ CKM matrix. 
This results a small deviation of $V_{ud}$ 
which is the most precisely measured by the experiments.     
There is the region where $\br{\ol{B}}{X_s\gamma}$ is deviated 
from the SM value at $\eps^{u}_{L_{2,3}} = 0$ at $\order{1\sigma}$, 
while the CKM elements are consistent with the current measurements.

\begin{table}[p]
 \centering 
\caption{\label{tab-BPs} 
Values of the observables at the benchmark points(BPs). 
}
\begin{tabular}[t]{c|cccc|c|cc} \hline 
 & BP1 & BP2 & BP3 & BP4 & SM & center & error \\  \hline\hline 
$\epsilon_2=\epsilon_3$ & 0.16 & 0.33 & 0.16 & 0.5 & 0 & 0 & 0 \\ 
\hline 
$\mathrm{BR}(\ol{B}\to X_s\gamma)$$\times 10^{4}$ & 4.125 & 3.919 & \
4.118 & 3.210 & 3.34 & 3.49 & 0.19 \\ 
$A_{\mathrm{CP}}$$\times 10^{5}$ & 8.38 & -0.998 & 9.15 & -1.77 & 0 & \
3690. & 2760. \\ \hline
$A_c$ & 0.668 & 0.668 & 0.668 & 0.667 & 0.668 & 0.67 & 0.027 \\ 
$A_b$ & 0.936 & 0.936 & 0.936 & 0.936 & 0.936 & 0.923 & 0.02 \\ 
$F_0$ & 0.697 & 0.697 & 0.697 & 0.697 & 0.697 & 0.684 & 0.0149 \\ 
$F_L$ & 0.3027 & 0.3027 & 0.3027 & 0.3027 & 0.3027 & 0.318 & 0.009 \
\\ 
$F_R$ & 0 & 0 & 0 & 0 & 0 & -0.002 & 0.0141 \\ 
$\mathrm{BR}(t\to cZ)$$\times 10^{4}$ & 0 & 0 & 0 & 0.019 & 0 & 0 \
& 1.2 \\ 
$\mathrm{BR}(t\to cH)$$\times 10^{4}$ & 0 & 0 & 0 & 0.0052 & 0 & 0 \
& 3.4 \\ 
$\mathrm{BR}(t\to c\gamma)$$\times 10^{12}$ & 0.0185 & 0.0185 & \
0.694 & 124. & 0 & 0 & $10^7$ \\ 
$R_{B_s}$ [\%] & 2.4 & 0.008 & 3.0 & 0 & 0 & 0 & 3.5 \\ 
 \hline \hline 
$m_u$ [MeV] & 1.2884 & 1.2884 & 1.2884 & 1.2884 & 1.2884 & \
1.2884 & 0.52231 \\ 
$m_c$ [GeV] & 0.627 & 0.627 & 0.627 & 0.627 & 0.627 & 0.627 & 0.0192 \\ 
$m_t$ [GeV] & 171.68 & 171.68 & 171.68 & 171.68 & 171.68 & 171.68 & 1.5147 \
\\ 
$m_d$ [MeV] & 2.7508 & 2.7508 & 2.7508 & 2.7508 & 2.7508 & \
2.7508 & 0.40044 \\ 
$m_s$ [MeV] & 54.32 & 54.32 & 54.32 & 54.32 & 54.32 & 54.32 \
& 2.96 \\ 
$m_b$ [GeV] & 2.85 & 2.85 & 2.85 & 2.85  & 2.85 & 2.85 & 0.0261 \\ 
\hline 
$|V_{ud}|$ & 0.97432 & 0.97432 & 0.97432 & 0.97432 & 0.97435 & \
0.97367 & 0.00032 \\ 
$|V_{us}|$ & 0.22515 & 0.22515 & 0.22515 & 0.22512 & 0.22501 & \
0.22431 & 0.00085 \\ 
$|V_{ub}|$$\times 10^{3}$ & 3.7229 & 3.7236 & 3.7271 & 3.7648 & 3.732 
& 3.82 & 0.2 \\ 
$|V_{cd}|$ & 0.225 & 0.225 & 0.225 & 0.2248 
& 0.22487 & 0.221 & 0.004 \\ 
$|V_{cs}|$ & 0.9734 & 0.9735 & 0.9734 & 0.9726 & 0.97349 & 0.975 & \
0.006 \\ 
$|V_{cb}|$$\times 10^{2}$ & 4.174 & 4.176 & 4.16 & 4.032 & 4.183 & \
4.11 & 0.12 \\ 
$|V_{td}|$$\times 10^{3}$ & 8.6711 & 8.6703 & 8.6772 & 8.7327 & 8.58 & \
8.6 & 0.2 \\ 
$|V_{ts}|$$\times 10^{2}$ & 4.1006 & 4.1001 & 4.1054 & 4.1484 & 4.111 \
& 4.15 & 0.09 \\ 
$|V_{tb}|$ & 0.999 & 0.999 & 0.999 & 0.998 & 0.999118 & 1.01 & 0.027 \\ 
\hline 
$\alpha$ & 1.57 & 1.57 & 1.57 & 1.58 & - & 1.47 & 0.0785 \\ 
$\sin 2\beta$ & 0.712 & 0.712 & 0.712 & 0.712 & - & 0.709 & 0.011 \
\\ 
$\gamma$ & 1.17 & 1.17 & 1.17 & 1.16 & - & 1.15 & 0.0524 \\
\hline 
\end{tabular}
\end{table}

Figure~\ref{fig_ACP} 
shows $\Delta A_{\CP}$ when the phase of the Yukawa coupling $\ola_d$ is rotated. 
This phase can be considered as those of certain combinations 
of $\eps^d_{X_i}$ with $i=2,3$ and $X=L,R,V$. 
We see that $\Delta A_{\CP}$ is at most $\order{0.1\%}$ 
in the cases (1) and (3), 
and it is $\order{0.01\%}$ in the case (2) 
due to the suppression by $\abs{C_7^\pr/C_7}^2$, see Eq.~\eqref{eq-ACP}.   
Thus the deviation is an order of magnitude smaller 
than the current limits.

Finally, we summarize the values of the observables 
at the benchmark points (BPs) in Table~\ref{tab-BPs}. 
The BPs are defined in the cases (1), (2), (3) and (4)  
with $\eps^q_{X_2} = \eps^q_{Y_3} = 0.16, 0.33, 0.16$ and $0.5$, 
respectively, where $(q,X,Y) = (d,L,R), (d,R,L), (d,V,V)$ and $(u,L,L)$.  
Note that the quark masses and the CKM observables 
are obtained after the fitting of the Yukawa couplings to the measured values. 
We see that the BPs are consistent with the SM predictions and experimental values  
except for $b\to s\gamma$.  
Thus, we show explicitly that $b \to s \gamma$ 
can give the most stringent bound on the vector-like fourth family quarks, 
especially due to the chiral enhancement effects.

\section{Summary and discussions} 
\label{sec-concl}

In this work, we demonstrated that a vector-like fourth family of quarks
induces a genuine chiral enhancement in $b\to s\gamma$,
which is absent in the SM.
The contributions to the Wilson coefficients $C_7$ and $C_7^\pr$
are approximately given by Eq.~\eqref{eq-C7app}
and are proportional to the Yukawa coupling constant $\ola_d$.
The factor $\ola_d v_H/m_b$ enhances the vector-like quark contributions,
leading to sizable deviations in $\br{\ol{B}}{X_s\gamma}$ from the SM prediction.

The mixing between the three chiral generations and the vector-like quarks 
induces the flavor violations of the $Z$ and Higgs bosons, 
as well as the deviation of the CKM matrix. 
Nevertheless, $\br{\ol{B}}{X_s\gamma}$ 
can be the most sensitive observables to the fourth family vector-like quarks. 
We examined the tree-level observables including the neutral meson mixing. 
The vector-like quark contributions 
are always subdominant compared with the SM ones, 
as long as the quark masses and CKM elements are well fit to the SM prediction.

This work focuses on the chiral enhancement of the vector-like quarks 
on the Wilson coefficients $C_7$ and $C_7^\pr$ 
through the triangle diagrams shown in Fig.~\ref{fig-FDs}. 
We expect that there are similar effects in other processes at loop-level. 
For instance, the box diagrams will contribute to the neutral meson mixing, 
and the $Z$ penguin diagrams will contribute to $K\to \pi \nu \nu$. 
The detailed studies of these processes are our future works.

\section*{Acknowledgement}

We thank S.~Iguro for useful discussions. 
This work is supported by the JSPS KAKENHI Grant Numbers 25K00222 (J.K.) and 24K07031 (Y.O.).

\appendix

\section{Analytical evaluation of mixing and couplings} 
\label{app-analytical} 

We derive the approximate analytical formula of the coupling matrices 
in the flavor space. 
We diagonalize the quark mass matrices, 
\begin{align}
 \Mcal_u = 
\begin{pmatrix}
 y^u_{i}\delta_{ij} v_H      & \eps^u_{R_i} v_H & 0_i \\
 \eps^u_{L_j} v_H  & \la_u v_H & m_U \\ 
 0_j & m_Q & \ola_u v_H
\end{pmatrix}, 
\quad 
\Mcal_d = 
\begin{pmatrix}
 y^d_{ij} v_H      & \eps^d_{R_i} v_H & 0_i \\
 \eps^d_{L_j} v_H  & \la_d v_H & m_D \\ 
 0_j & m_Q & \ola_d v_H
\end{pmatrix}, 
\end{align}
for $\eps^q_{X_i}, v_H/m_\VLQ \ll 1$ with $m_\VLQ \sim m_Q, m_U, m_D$.

First, we diagonalize the mass matrices at $\eps^{q}_{X_i} = 0$, 
so that the matrices are block diagonal.   
Each block can be diagonalized as  
\begin{align}
\label{eq-UMUbar}
\left(U^0_{q_R}\right)^\dag \Mcal_q U^0_{q_L}
=&\  
\begin{pmatrix}
 y_i^q \delta_{ij} v_H & c_{q_L} \teps^q_{R_i} v_H 
                       & s_{q_L} \teps^q_{R_i} v_H \\ 
 s_{q_R} \teps^q_{L_j} v_H & m_{q_4} & 0 \\ 
 c_{q_R} \teps^q_{L_j} v_H & 0 & m_{q_5}    
\end{pmatrix},
\end{align}
where 
\begin{align}
 U^0_{u_X} = \id{3} \otimes w_{u_X}, 
 \quad 
 U^0_{d_X} = u_X \otimes w_{d_X}. 
\end{align}
Here, $\otimes$ means that the matrix is block diagonalized as  
$(3\times 3)\otimes (2\times 2)$. 
The unitary matrices $u_{X}$ and $w_{q_X}$ 
are the diagonalization matrices for each block, i.e. 
\begin{align}
 u_R^\dag y_d u_L =&\ \mathrm{diag}(y^d_1, y^d_2, y^d_3), 
\\ \notag 
 w_{u_R}^\dag 
\begin{pmatrix}
 \la_u v_H & m_U \\ m_Q & \ola_u v_H
\end{pmatrix} 
 w_{u_L} 
 =&\ \mathrm{diag}(m_{u_4}, m_{u_5}), 
\quad  
 w_{d_R}^\dag 
\begin{pmatrix}
 \la_d v_H & m_D \\ m_Q & \ola_d v_H
\end{pmatrix} 
 w_{u_L} 
 = \mathrm{diag}(m_{d_4}, m_{d_5}). 
\end{align}
We parametrize $w_{q_X}$ as 
\begin{align}
 w_{q_L} = 
\begin{pmatrix}
 c_{q_L}& s_{q_L} \\ -s_{q_L}^* & c_{q_L}
\end{pmatrix}, 
\quad 
 w_{q_R} = 
\begin{pmatrix}
 s_{q_R}^* & c_{q_R} \\ c_{q_R} & -s_{q_R}
\end{pmatrix}, 
\end{align}
where $c_{q_X}$ and $s_{q_X}$ satisfy $c_{q_X}^2 + \abs{s_{q_X}}^2 = 1$. 
Here, the former is real while the latter is complex. 
In this notation, 
\begin{align}
 m_{q_4} \simeq m_Q + \order{v_H}, \quad 
 m_{u_5} \simeq m_{U} + \order{v_H}, \quad  
 m_{d_5} \simeq m_{D} + \order{v_H},   
\end{align}
with $m_{R_u} := m_U$ and $m_{R_d} := m_D$ 
and the mixing angles are given by  
\begin{align}
 c_{q_X} \sim 1, 
\quad 
 s_{q_L} \sim -\frac{\ola_q m_Q + \la_q^* m_{R_q}}{m_Q+m_{R_q}} 
               \frac{v_H}{m_Q-m_{R_q}}, 
\quad 
 s_{q_R} \sim  \frac{\la_q^* m_Q + \ola_q m_{R_q}}{m_Q+m_{R_q}} 
               \frac{v_H}{m_Q-m_{R_q}},  
\end{align}
up to $\order{v_H}$. 
After the diagonalization of each block, 
the off-diagonal blocks are proportional to 
\begin{align}
\teps^u_{L_i} := \eps^u_{L_i}, 
\quad 
\teps^u_{R_i} := \eps^u_{R_i}, 
\quad 
\teps^d_{L_i} := \eps^d_{L_k} u_L^{ki}, 
\quad  
\teps^d_{R_i} := (u_R^\dag)_{ik} \eps^d_{R_k}.  
\end{align}
These will be a source for flavor violations in the mass basis.

The off-diagonal blocks at $\order{\eps^q_{X}}$ in Eq.~\eqref{eq-UMUbar}   
can be approximately diagonalized as 
\begin{align}
\label{eq-MdiagApp}
 \left(U^0_{q_R}U^1_{q_R}\right)^\dag 
 \Mcal_q
 U^0_{q_L} U^1_{q_L} 
= 
\begin{pmatrix}
 v_H \left(y_i^q \delta_{ij}
  -\dfrac{m_{q_a}}{v_H} \Ecal^{q_R*}_{ai} \Ecal^{q_L}_{aj}\right) 
 & v_H y_i \Ecal^{q_L*}_{4i}   & v_H y_i \Ecal^{q_L*}_{4i}   
\\ 
  v_H y_j  \Ecal^{q_R}_{4j} & m_{q_4} + \order{v_H \Ecal^2} & \order{v_H \Ecal^2}
\\ 
  v_H y_j  \Ecal^{q_R}_{5j} & \order{v_H \Ecal^2} & m_{q_5} + \order{v_H \Ecal^2}
\end{pmatrix}
\end{align}
where the sum over $a=4,5$ is implicit. 
the unitary matrices are defined as 
\begin{align}
 U^1_{q_{X}} =  
\begin{pmatrix}
 \delta_{ij} - \dfrac{1}{2} \Ecal^{q_X*}_{ai} \Ecal^{q_X}_{aj} 
 & \Ecal^{q_X*}_{4i} & \Ecal^{q_X*}_{5i}  \\ 
 - \Ecal^{q_X}_{4i} & 1 + \order{\Ecal^2} & \order{\Ecal^2} \\ 
 - \Ecal^{q_X}_{5i} & \order{\Ecal^2} & 1+\order{\Ecal^2}   \\ 
\end{pmatrix}, 
\end{align} 
with $X=L,R$ and $q=u,d$. 
Here, we define  
\begin{align}
 \Ecal^{q_L}_{4i} := s_{q_R} \teps^q_{L_i} \frac{v_H}{m_{q_4}}, 
\quad  
 \Ecal^{q_L}_{5i} := c_{q_R} \teps^q_{L_i} \frac{v_H}{m_{q_5}},
\quad 
 \Ecal^{q_R}_{4i} := c_{q_L}   \teps^{q*}_{R_i} \frac{v_H}{m_{q_4}}, 
\quad  
 \Ecal^{q_R}_{5i} := s_{q_L}^* \teps^{q*}_{R_i} \frac{v_H}{m_{q_5}}. 
\end{align}
Equation~\eqref{eq-MdiagApp} is diagonalized 
up to $\order{m_\VLQ \Ecal^2, y_i v_H \Ecal}$, 
and thus we will evaluate the coupling matrices in the mass basis, 
rotated by 
\begin{align}
 U_{q_X} \simeq U_{q_X}^0 U_{q_X}^1.  
\end{align}

Finally, the flavor structure of the $W$ boson conplings are given by
\begin{align}
&\  U^\dag_{u_L} H_L U_{d_L} 
\\ \notag 
&\ = 
\left(
\begin{matrix}
   u_L^{ij} + \delta_L u_{ij} 
\\ 
 \Ecal^{u_L}_{4k} u_{kj} 
  - c_{u_L} \left(c_{d_L} \Ecal^{d_L}_{4j} + s_{d_L} \Ecal^{d_L}_{5j} \right)
\\ 
  \Ecal^{u_L}_{5k} u_{kj} 
  - s_{u_L}^* \left(c_{d_L} \Ecal^{d_L}_{4j} + s_{d_L} \Ecal^{d_L}_{5j} \right)
\end{matrix}
\right.
\\ \notag 
& \hspace{4.0cm}
\left. 
\begin{matrix}
    u_{i\ell} \Ecal^{d_L*}_{4\ell} 
   -c_{d_L}\left(c_{u_L} \Ecal^{u_L*}_{4i}+ s^*_{u_L}  \Ecal^{u_L*}_{5i} \right) 
& 
 u_{i\ell} \Ecal^{d_L*}_{5\ell} 
   -s_{d_L}\left(c_{u_L} \Ecal^{u_L*}_{4i}+ s^*_{u_L}  \Ecal^{u_L*}_{5i} \right)  
\\
 c_{u_L} c_{d_L} & c_{u_{L}} s_{d_L} 
\\ 
s_{u_L}^* c_{d_L} & s_{u_{L}}^* s_{d_L} 
\end{matrix}
\right), 
\\ 
& U^\dag_{u_R} H_R U_{d_R} 
\\ \notag 
&\ = 
\begin{pmatrix}
 \delta_R u_{ij}& 
   -c_{d_R}\left(c_{u_R} \Ecal^{u_R*}_{4i}-s_{u_R}^* \Ecal^{u_R*}_{5i} \right)
&
    s_{d_R}\left(c_{u_R} \Ecal^{u_R*}_{4i}-s_{u_R}^* \Ecal^{u_R*}_{5i} \right)
\\ 
 -c_{u_R}\left(c_{d_R} \Ecal^{d_R}_{4j} - s_{d_R} \Ecal^{d_R}_{5j}  \right) 
 & c_{u_R}c_{d_R} & -c_{u_R} s_{d_R} 
\\ 
 s_{u_R}^* \left(c_{d_R} \Ecal^{d_R}_{4j} - s_{d_R} \Ecal^{d_R}_{5j}  \right) 
 & -s_{u_R}^* c_{d_R} & s_{u_R}^* s_{d_R} 
\end{pmatrix}, 
\end{align}
where the non-unitary parts of the $3\times 3$ block are given by 
\begin{align}
 \delta_L u_{ij} 
=&\  -\frac{1}{2} \sum_{a=4,5} \left(\Ecal^{u_L*}_{ai} \Ecal^{u_L}_{ak} u_L^{kj} 
           + u_L^{i\ell} \Ecal^{d_L*}_{a\ell} \Ecal^{d_L}_{aj}  \right)
 + \left(c_{u_L}\Ecal^{u_L}_{4i}+s_{u_L}\Ecal^{u_L}_{5i}\right)^* 
   \left(c_{d_L}\Ecal^{d_L}_{4j}+s_{d_L}\Ecal^{d_L}_{5j}\right), 
\notag \\
 \delta_R u_{ij} 
=&\  
   \left(c_{u_R}\Ecal^{u_R}_{4i}-s_{u_R}\Ecal^{u_R}_{5i}\right)^*
   \left(c_{d_R}\Ecal^{d_R}_{4j}-s_{d_R}\Ecal^{d_R}_{5j}\right).       
\end{align}
Those for the $Z$-boson couplings are given by 
\begin{align}
 U_{q_L}^\dag H_L U_{q_L} = &\ 
 \begin{pmatrix}
  \delta_{ij} + \Delta^{q_L}_{ij} & 
  s^*_{q_L}\left( s_{q_L} \Ecal^{q_L*}_{4i} - c_{q_L} \Ecal^{q_L*}_{5i} \right) & 
  - c_{q_L}\left( s_{q_L} \Ecal^{q_L*}_{4i} - c_{q_L} \Ecal^{q_L*}_{5i} \right) 
\\ 
s_{q_L}\left( s_{q_L}^* \Ecal^{q_L}_{4i} - c_{q_L} \Ecal^{q_L}_{5i} \right) & 
c_{q_L}^2 & c_{q_L}s_{q_L}  
\\
-c_{q_L}\left( s_{q_L}^* \Ecal^{q_L}_{4i} - c_{q_L} \Ecal^{q_L}_{5i} \right) 
& c_{q_L}s_{q_L}^*  & \abs{s_{q_L}}^2 
 \end{pmatrix},
\\ \notag 
 U^\dag_{q_R} H_R U_{q_R}  = &\ 
\begin{pmatrix}
 \Delta^{q_R}_{ij} & 
   -c_{q_R}\left(c_{q_R} \Ecal^{q_R*}_{4i}-s_{q_R}^* \Ecal^{q_R*}_{5i} \right)
&
    s_{q_R}\left(c_{q_R} \Ecal^{q_R*}_{4i}-s_{q_R}^* \Ecal^{q_R*}_{5i} \right)
\\ 
  -c_{q_R}\left(c_{q_R} \Ecal^{q_R}_{4j} - s_{q_R} \Ecal^{q_R}_{5j}  \right) 
 & c_{q_R}^2 & -c_{q_R} s_{q_R} 
\\ 
   s_{q_R}^* \left(c_{q_R} \Ecal^{q_R}_{4j} - s_{q_R} \Ecal^{q_R}_{5j}  \right) 
& -s_{q_R}^* c_{q_R} & \abs{s_{q_R}}^2 
\end{pmatrix}, 
\end{align}
for $q=u,d$, where the corrections to the SM parts read as 
\begin{align}
 \Delta^{q_L}_{ij}  =  
 \begin{pmatrix}
  \Ecal^{q_L}_{4i} \\   \Ecal^{q_L}_{5i} 
 \end{pmatrix}^\dagger 
\begin{pmatrix}
-\abs{s_{q_L}}^2  & c_{q_L}s_{q_L} \\ s_{q_L}^*c_{q_L} & -c_{q_L}^2 
\end{pmatrix}
 \begin{pmatrix}
  \Ecal^{q_L}_{4j} \\ \Ecal^{q_L}_{5j} 
 \end{pmatrix},
\quad 
 \Delta^{q_R}_{ij}  =  
 \begin{pmatrix}
  \Ecal^{q_R}_{4i} \\ \Ecal^{q_R}_{5i} 
 \end{pmatrix}^\dagger 
\begin{pmatrix}
 c_{q_R}^2 & -c_{q_R}s_{q_R} \\ -s_{q_R}^*c_{q_R} & \abs{s_{q_R}}^2 
\end{pmatrix}
 \begin{pmatrix}
  \Ecal^{q_R}_{4j} \\ \Ecal^{q_R}_{5j} 
 \end{pmatrix}. 
\end{align}

The Higgs Yukawa coupling matrix is given by 
\begin{align}
 U^\dag_{q_R} Y_q U_{q_L} =&\ 
\left(
\begin{matrix}
 y_i^q \delta_{ij} -(m_{q_a}/v_H)\Ecal^{q_R*}_{ai}\Ecal^{q_L}_{aj} 
   + \delta y^q_{ij} 
\\ 
   \Ecal^{q_R}_{4j} y^q_j-\lambda_{4a}^q \Ecal^{q_L}_{aj}+(m_{q_4}/v_H)\Ecal^{q_L}_{4j}  
\\ 
   \Ecal^{q_R}_{5j} y^q_j-\lambda_{5a}^q \Ecal^{q_L}_{aj}+(m_{q_5}/v_H)\Ecal^{q_L}_{5j} 
\end{matrix}
\right.
\\ \notag 
& \hspace{1.8cm}  \left. 
\begin{matrix}
 y_i\Ecal^{q_L*}_{4i}-\lambda^q_{a4} \Ecal^{q_R*}_{ai}+(m_{q_4}/v_H)\Ecal^{q_R*}_{4i} 
&
 y_i\Ecal^{q_L*}_{5i}-\lambda^q_{a5} \Ecal^{q_R*}_{ai}+(m_{q_5}/v_H)\Ecal^{q_R*}_{5i}
\\ 
\lambda_{44}^q & \lambda_{45}^q \\ 
\lambda_{54}^q & \lambda_{55}^q 
\end{matrix}
\right), 
\end{align}
where 
\begin{align}
 \begin{pmatrix}
  \la^q_{44} & \la^q_{45} \\   \la^q_{54} & \la^q_{55} 
 \end{pmatrix}
:= 
\begin{pmatrix}
 s_{q_R} c_{q_L} \la_q - c_{q_R}s_{q_L}^* \ol{\lambda}_q 
& 
 s_{q_R} s_{q_L} \la_q + c_{q_R}c_{q_L} \ol{\lambda}_q 
\\ 
 c_{q_R} c_{q_L} \la_q + s_{q_R}^*s_{q_L}^* \ol{\lambda}_q 
& 
 c_{q_R} s_{q_L} \la_q - s_{q_R}^*c_{q_L} \ol{\lambda}_q 
\end{pmatrix},
\end{align}
and $\order{\Ecal^2}$ is omitted except for $\delta y_{ij}$. 
Here, the sum over $a,b=4,5$ are implicit. 
The tree-level flavor violations will be induced by   
\begin{align}
 \delta y^q_{ij} = &\ 
   - \frac{m_{q_a}}{v_H} \Ecal^{q_R*}_{ai} \Ecal^{q_L}_{aj} 
   -\frac{1}{2} \left(y_i \Ecal^{q_L*}_{ai} \Ecal^{q_L}_{aj}   
   + \Ecal^{q_R*}_{ai} \Ecal^{q_R}_{aj} y_j \right)
   + \Ecal^{q_R*}_{ai}\lambda_{ab}^q \Ecal^{q_L}_{bj}. 
\end{align}
We separate the Yukawa matrix in the SM block 
whose form is the same as the mass matrix in Eq.~\eqref{eq-MdiagApp}.

To simplify the formulas, we further assume $v_H/(m_Q-m_{U,D}) \ll 1$, so that 
\begin{align}
\abs{ s_{q_X}} \ll 1. 
\end{align} 
In the main text, we keep the leading part in the expansion in $s_{q_X}$. 
This approximation is not hold for the degenerate case, 
but similar cancellation would occur by the degeneracy of the masses.   
In addition, we neglect the SM Yukawa couplings 
and the mixing angles in the CKM matrix, i.e. 
$y_i \ll 1$.

Using the above results, we obtain the leading contribution 
to $C_7$ and $C_7^\pr$ through the vector-like quark loops. 
The $W$ boson loop contributions are given by 
\begin{align}
\label{eq-C7appW}
 C_{7,W}^Q(\mu_Q) 
\simeq&\ 
 -\frac{D^{\prime}_W(w_{U})}{w_U} \frac{\heps^{u*}_{L_2}\heps^u_{L_3}}{g^2 \la^W_{3LL}}
\\ \notag 
&\
 +\frac{v_H}{g^2m_b} \left\{
  \left(\la_u + \ol{\la}_u^* \sqrt{\frac{w_U}{w_Q}} \right) 
  \frac{D^{\prime\prime}_W(w_Q)-D^{\prime\prime}_W(w_U)}{w_Q-w_U}
  \frac{\heps^{u*}_{L_2} \teps^{d*}_{R_3}}{\la^W_{3LL}}  
 +\ol{\la}_d^* \frac{D^{\prime\prime}_W(w_Q)}{\sqrt{w_Qw_D}}  
  \frac{\teps^{d*}_{L_2}\teps^{d*}_{R_3}}{\la^W_{3LL}}
  \right\},
\\
C_{7,W}^{Q\prime}(\mu_Q) 
\simeq &\ -
       \frac{D^{\prime}_W(w_Q)}{w_Q} 
       \frac{\teps^{d}_{R_2}\teps^{d*}_{R_3}}{g^2 \la^W_{3LL}} 
\\ \notag 
&\  
  + \frac{v_H}{g^2 m_b} \left\{
  \left( \la_u^* + \ol{\la}_u \sqrt{\frac{w_U}{w_Q}} \right) 
  \frac{D^{\prime\prime}_W(w_Q)-D^{\prime\prime}_W(w_U)}{w_Q-w_U}
   \frac{\teps^{d}_{R_2} \heps^{u}_{L_3}}{\la^W_{3LL}} 
 +\ol{\la}_d \frac{D^{\prime\prime}_W(w_Q) }{\sqrt{w_Qw_D}} 
  \frac{\teps^{d}_{R_2}\teps^{d}_{L_3}}{\la^W_{3LL}}
  \right\},  
\end{align}
where $\heps^u_{{L_i}} := \eps^u_{L_k} u^L_{ki}$ 
and $w_F := m_F^2/m_W^2$ with $F=Q,U,D$.  
The $Z$ and Higgs boson contributions are respectively given by 
\begin{align}
C_{7,Z}^{Q}(\mu_Q) \simeq&\ 
 \frac{c_W^2}{2} Q_d \Biggl[
- \frac{E_V^\pr(z_D)}{z_D} \frac{\teps^{d*}_{L_2}\teps^d_{L_3}}{g^2\la^W_{3LL}}
\\ \notag 
&\  + \frac{v_H}{g^2m_b(z_Q-z_D)}  
 \left\{ 
 \left(\la_d+\sqrt{\frac{z_Q}{z_D}}\ola^*_{d}\right) E^{\pr\pr}_V(z_Q) 
-\left(\la_d+\sqrt{\frac{z_D}{z_Q}}\ola^*_{d}\right) E^{\pr\pr}_V(z_D) 
  \right\}
 \frac{\teps^{d*}_{L_2}\teps^{d*}_{R_3}}{\la^W_{3LL}} 
 \Biggr], 
 \\ \notag 
C_{7,Z}^{Q\prime} \simeq &\ 
 \frac{c_W^2}{2} Q_d \Biggl[
- \frac{E_V^\pr(z_Q)}{z_Q} \frac{\teps^{d}_{R_2}\teps^{d*}_{R_3}}{g^2\la^W_{3LL}}
\\ \notag 
&\  + \frac{v_H}{g^2m_b(z_Q-z_D)}  
 \left\{ 
 \left(\la_d^*+\sqrt{\frac{z_Q}{z_D}}\ola_{d}\right) E^{\pr\pr}_V(z_Q) 
-\left(\la_d^*+\sqrt{\frac{z_D}{z_Q}}\ola_{d}\right) E^{\pr\pr}_V(z_D) 
  \right\}
 \frac{\teps^{d}_{R_2}\teps^{d}_{L_3}}{\la^W_{3LL}} 
 \Biggr], 
\end{align}
and 
\begin{align}
C^Q_{7,H} 
=&\ 
 -\frac{Q_d m_W^2}{2g^2 m_H^2} 
\Biggl[
  E^\pr_S(h_D) \frac{\teps^{d*}_{L_2}\teps^d_{L_3}}{\la^W_{3LL}} 
+ \frac{v_H}{m_b(h_Q-h_D)} 
\Biggl(
 \left\{
 \la_d+\left(-\sqrt{\frac{h_Q}{h_D}}+2\sqrt{\frac{h_D}{h_Q}}\right)\ola_d^*
 \right\} h_Q E_S^{\pr\pr}(h_Q)
\\ \notag 
&\ \hspace{4.0cm}
- 
 \left\{
 \la_d+\left(-\sqrt{\frac{h_D}{h_Q}}+2\sqrt{\frac{h_Q}{h_D}}\right)\ola_d^*
 \right\} h_D E_S^{\pr\pr}(h_D)
\Biggr)
\frac{\teps^{d*}_{L_2} \teps^{d*}_{R_3}}{\la^W_{3LL}} 
\Biggr],
\\ 
C^{Q\prime}_{7,H} 
\simeq &\  
-\frac{Q_d m_W^2}{2g^2 m_H^2}
\Biggl[
  E^\pr_S(h_Q) \frac{\teps^{d}_{R_2}\teps^{d*}_{R_3}}{\la^W_{3LL}} 
+ \frac{v_H}{m_b(h_Q-h_D)} 
\Biggl(
 \left\{
 \la_d^*+\left(-\sqrt{\frac{h_Q}{h_D}}+2\sqrt{\frac{h_D}{h_Q}}\right)\ola_d
 \right\} h_Q E_S^{\pr\pr}(h_Q)
\\ \notag 
&\ \hspace{4.0cm}
- 
 \left\{
 \la_d^*+\left(-\sqrt{\frac{h_D}{h_Q}}+2\sqrt{\frac{h_Q}{h_D}}\right)\ola_d 
 \right\} h_D E_S^{\pr\pr}(h_D)
\Biggr)
\frac{\teps^{d}_{R_2} \teps^{d}_{L_3}}{\la^W_{3LL}} 
\Biggr],
\end{align}
where $z_F = m_F^2/m_Z^2$ and $h_F = m_F^2/m_H^2$. 
We can show Eq.~\eqref{eq-C7app} by inserting the asymptotic formula 
of the loop functions.

\section{Evaluation of numerical coefficients}  
\label{app-coes}

We summarize our evaluation 
of the numerical constants to calculate 
$\ol{B}\to X_s\gamma$ and $B_s$-$\ol{B}_s$ mixing. 

\subsection{\texorpdfstring{$\overline{B}\to X_s\gamma$}{Bbar -> Xs gamma}}
\label{app-BsGamma}

\subsubsection{Anomalous dimensions}

To include the scale dependence of the Wilson coefficients, 
we solve the evolution equation 
\begin{align}
 \frac{dC_i(\mu)}{d\ln\mu} = \frac{\alpha_s(\mu)}{4\pi} \sum_j C_j(\mu) \gamma_{ji},  
\end{align} 
where the coefficients are defined in Eqs.~\eqref{eq-Q18},~\eqref{eq-Q910} 
and~\eqref{eq-Qnn}.

For the upper-left $6\times 6$ block, 
the anomalous dimension matrix is given by~\cite{Vainshtein:1976xii} 
\begin{align}
 \gamma_{6\times 6} = 
 \begin{pmatrix}
  -2 & 6 &  0    & 0 & 0 & 0 \\ 
 6 & -2  & -2/9  & 2/3 & -2/9 & 2/3 \\ 
 0 & 0   & -22/9 & 22/3& -4/9 & 4/3 \\ 
 0 & 0   & 6-2f/9& -2+2f/3 & -2f/9 &   2f/3 \\
 0 & 0   & 0 & 0 & 2 & -6 \\ 
 0 & 0   & -2f/9 & 2f/3 & -2f/9 & -16 + 2f/3  
 \end{pmatrix}, 
\end{align}
where $[\gamma_{6\times6}]_{ij} := \gamma_{ij}$ for $i,j = 1,2,\cdots,6$ 
and $f$ is the number of quark flavors. 
The elements involving $Q_{7,8}$ are given by~\cite{Ciuchini:1993ks,Ciuchini:1993fk}  
\begin{align}
\vec{\gamma}_7 =&\ 
\begin{pmatrix}
0& \dfrac{416}{81}& -\dfrac{464}{81}& -\dfrac{232}{81}f_d +\dfrac{416}{81} f_u & 
\dfrac{32}{9} & \dfrac{200}{81}f_d-\dfrac{448}{81} f_u 
& \dfrac{32}{3} & -\dfrac{32}{9}
\end{pmatrix}, 
\\ \notag 
\vec{\gamma}_8 =&\ 
\begin{pmatrix}
3 & \dfrac{70}{27} & \dfrac{140}{27}+3f & 6+\dfrac{70}{27}f & -\dfrac{14}{3}-3f 
&-4-\dfrac{114}{27}f & 0 & \dfrac{28}{3} 
\end{pmatrix}, 
\end{align}
where $[\vec{\gamma}_a]_i := \gamma_{ia}$ for $i=1,2,\cdots,8$ and $a=7,8$. 
Here, $f_u$ ($f_d$) is the number of flavor of the up-type (down-type) quarks. 
Those involving the left-right operators $Q_{9}$ and $Q_{10}$ 
are given by~\cite{Cho:1993zb}   
\begin{align}
\vec{\gamma}_9 =&\ 
\begin{pmatrix}
 \dfrac{160}{3} & -4 & -16 & 0
\end{pmatrix},
\quad 
\vec{\gamma}_{10} =
\begin{pmatrix}
 \dfrac{64}{9} & \dfrac{8}{3} & -6 & 2 
\end{pmatrix}, 
\end{align}
where $[\vec{\gamma}_b]_j := \gamma_{bj}$ for $j=7,8,9,10$ and $b=9,10$. 
The other elements in the full $10\times 10$ matrix are vanishing.

The anomalous dimensions involving the neutral current operators 
are given as follows~\cite{Buras:1993dy,Buchalla:1989we,Buras:2011zb}. 
Those between the neutral current operators $(Q^f_{1}(L,X), Q^f_{2}(L,X))$ 
are given in the block diagonalized forms, 
\begin{align}
 \gamma_{nn}(L,L) = 
\begin{pmatrix}
 -2 & 6 \\ 6 & -2
\end{pmatrix}, 
\quad 
 \gamma_{nn}(L,R) = 
\begin{pmatrix}
 -16 & 0 \\ -6 & 2
\end{pmatrix}, 
\end{align} 
which are the same as the charged current operators. 
The off-diagonal blocks of $(Q_{7}, Q_{8})$ and   
$(Q_1^{u,c,t}, Q_2^{u,c,t}, 
  Q_1^d, Q_2^d, Q_1^{s,b}, Q_2^{s,b}, \hat{Q}_1^d, \hat{Q}^d_2)$ 
are given by 
\begin{align}
 \gamma^T_{nD}(L,L) =&\  
\begin{pmatrix}
 \dfrac{416}{81} & 0 & -\dfrac{232}{81} & 0 & -\dfrac{232}{81} & -\dfrac{232}{81} 
 & 0 & -\dfrac{232}{81} \\ 
\\
 \dfrac{70}{27} & 3 & \dfrac{70}{27} & 3 & \dfrac{151}{27} & \dfrac{151}{27} & 3
 & \dfrac{70}{27} 
\end{pmatrix},
\\ \notag 
 \gamma^T_{nD}(L,R) =&\  
\begin{pmatrix}
- \dfrac{448}{81} & 0 & \dfrac{200}{81} & 0 & \dfrac{200}{81} & \dfrac{16}{9} 
 & -\dfrac{80}{3} & -\dfrac{32}{9} \\ 
\\
 -\dfrac{119}{27} & -3 & -\dfrac{119}{27} & -3 & -\dfrac{173}{27} & 
-\dfrac{16}{27} & -4 & \dfrac{8}{3} 
\end{pmatrix}.
\end{align}
All of the neutral current operators 
\begin{align}
 Q^{nn} = \left\{
 Q_1^{u,c,t,d,s,b}(L,L), Q_2^{s,b}(L,L), \hat{Q}^d_2(L,L), 
 Q_1^{u,c,t,d,s,b}(L,R) 
 \right\}, 
\end{align}
have the universal off-diagonal anomalous dimensions with the operators 
$(Q_3, Q_4, Q_5, Q_6)$ as 
\begin{align}
 \gamma_{nP} = 
\begin{pmatrix}
 -\dfrac{2}{9} & \dfrac{2}{3} &  -\dfrac{2}{9} & \dfrac{2}{3}   
\end{pmatrix}. 
\end{align}
The other elements involving $Q^{nn}$ are vanishing.

\subsubsection{Evaluation of \texorpdfstring{$\kappa$}{kappa} and \texorpdfstring{$\rho$}{rho} }

The solution for the evolution equation is given by 
\begin{align}
 C_i(\mu) =&\ U_{ij}(\mu,\mu_0) C_j(\mu_0),  
\\ \notag 
 U(\mu,\mu_0) =&\ V \left(\frac{\alpha(\mu)
                \left\{1+\beta_1 {\alpha(\mu_0)}/{(4\pi\beta_0)}\right\}}
               {\alpha({\mu_0})\left\{1+\beta_1 \alpha(\mu)/(4\pi\beta_0)\right\}}
               \right)^{-\gamma_D/(2\beta_0)} V^{-1},
\end{align}
where the transposed anomalous dimension is diagonalized by a matrix $V$ as  
\begin{align}
 \gamma_D = V^{-1} \gamma_0^T V. 
\end{align}
Here, we solved the 2-loop RGE of the strong coupling constant 
\begin{align}
 \frac{d\alpha}{d\ln \mu} = -2 \beta_0 \frac{\alpha^2}{4\pi} 
                            -2 \beta_1 \frac{\alpha^3}{16\pi^2}, 
\end{align}
with 
\begin{align}
 \beta_0 = 11-\frac{2}{3}f, \quad 
 \beta_1 = 102-\frac{38}{3}f. 
\end{align}
The solution of the RGE is given by 
\begin{align}
 \alpha(\mu) =- \frac{4\pi\beta_0}{(1+w(\mu))\beta_1}, 
\end{align}
with 
\begin{align}
 w(\mu) := \Wcal_{-1}\left( 
-\left(1+\frac{4\pi\beta_0}{\beta_1 \alpha(\mu_0)} \right)
      e^{-\left(1+\frac{4\pi\beta_0}{\beta_1 \alpha(\mu_0)}\right)} 
      \left(\frac{\mu_0}{\mu} \right)^{\frac{2\beta_0^2}{\beta_1}}
 \right), 
\end{align}
where $\Wcal_{-1}$ is the Lambert function 
satisfying $z = \Wcal_{-1}(z) e^{\Wcal_{-1}(z)}$ 
on the branch $\Wcal_{-1}(z) < -1$ with $ -e^{-1} < z < 0$.

\begin{table}[t]
\center
\caption{\label{tab-input}
Values of the inputs~\cite{ParticleDataGroup:2024cfk}.  
}
\begin{tabular}[t]{c|c} \hline
 input & value  \\ \hline\hline 
 $\alpha(m_Z)$ & 0.1180 (9)  \\
 $m_Z$ [GeV]   & 91.1880 (29)   \\ 
 $m_W$ [GeV]   & 80.3692 (133)   \\ 
 $m_t(m_t)$ [GeV] & 162.690 (6)  \\ 
\hline
\end{tabular}
\end{table}

\begin{table}[t]
\center 
\caption{\label{tab-kappa78}
Values of $\kappa_{7,8}$ at $\mu_b = 2.5~\GeV$ 
with various heavy mass scales $\mu_Q$.   
}
\begin{tabular}[t]{c|c|cccc} \hline 
$\mu_Q$ [GeV]    & $m_W$ & 200   & 1000  & 5000  & 10000  \\ 
\hline \hline 
$\kappa_7^{(Q)}$ & 0.600 & 0.550 & 0.482 & 0.451 & 0.412  \\ 
$\kappa_8^{(Q)}$ & 0.105 & 0.114 & 0.123 & 0.126 & 0.129  \\
\hline 
$\rho_8^{(Q)}$   & 0.640 & 0.593 & 0.528 & 0.499 & 0.461 \\
\hline 
\end{tabular}
\end{table}

We employ the values of the gauge coupling constant and masses 
shown in Table~\ref{tab-input}
to evaluate the RG coefficients in Eq.~\eqref{eq-DeltaC7}.  
The values of $\kappa_{7,8}^Q$ at various $\mu_Q$ are shown in Table~\ref{tab-kappa78}. 
The other coefficients relevant to our analysis are given by 
\begin{align}
\kappa_2 = -0.219, \quad \kappa_{10} = -0.686, 
\end{align}
\begin{align}
 \kappa_{LL}^{2,u} = \kappa_{LL}^{2,c} = 0.0251, \quad 
 \kappa_{LL}^{2,d}   = -0.0276, \quad 
 \kappa_{LL}^{2,s} = \kappa_{LL}^{2,b} = 0.0810, \quad 
 \hat{\kappa}_{LL}^{2,d} = 0.109, 
\end{align}
and 
\begin{align}
 \kappa_{LR}^{2,u} =&\  \kappa_{LR}^{2,c} = 0.0543, 
&\quad& 
 \kappa_{LR}^{2,d}   = -0.0089, 
&\quad & 
 \kappa_{LR}^{2,s} = \kappa_{LR}^{2,b} = -0.0655, 
&\quad & 
 \hat{\kappa}_{LR}^{2,d} = 0.333,  
\end{align}
where the scale is set to $\mu=m_Z$. 
These values can also be used for $C_7^\prime$ as the QCD is not chiral.

Similarly to $C_7$, we evaluate the RG coefficients on $C_8$, denoted by $\rho$, as  
\begin{align}
\label{eq-DeltaC7Rho}
\Delta C_8(\mu_b) =&\ 
       \rho_8^Q C^Q_8(\mu_Q) + \rho_8^q \Delta C^q_8(\mu_W) 
  \\ \notag 
 &\ \hspace{-1.5cm} 
+  \rho_2 \Delta C_2(\mu_W) + \rho_{10} C_{10}(\mu_W) 
       + \sum_{a=1,2}\sum_{X=L,R}\left(
                      \hat{\rho}_{LX}^{a,d} \hat{C}_2^d(\mu_Z; L,X) 
                   + \sum_f \rho_{LX}^{a,f} C_a^{f}(\mu_Z; L,X)  \right).
\end{align}
The values of $\rho_{8}^Q$ at several $\mu_Q$ 
are shown in Table~\ref{tab-kappa78}.  
The other coefficients are given by 
\begin{align}
\rho_2 = -0.0997, \quad \rho_{10} = -0.0673,  
\end{align}
\begin{align}
& \rho_{LL}^{2,u} = \rho_{LL}^{2,c} = \rho_{LL}^{2,d} =  -0.108, 
&\quad& 
 \rho_{LL}^{2,s} = \rho_{LL}^{2,b} = -0.208, 
&\quad&  
\hat{\rho}_{LL}^{2,d} = -0.0997,  
\end{align}
and 
\begin{align}
& \rho_{LR}^{1,u} = \rho_{LR}^{1,c} = 0.258, 
&\quad& 
 \rho_{LR}^{1,s} = \rho_{LR}^{1,b} = 0.381, 
&\quad& 
 \hat{\rho}_{LR}^{1,d} = 0.246, 
\\ \notag 
& \rho_{LR}^{2,u} = \rho_{LR}^{2,c} = 0.146, 
&\quad& 
 \rho_{LR}^{2,s} = \rho_{LR}^{2,b} = 0.250, 
&\quad& 
 \hat{\rho}_{LR}^{2,d} = -0.0673,   
\end{align}
at $\mu= m_Z$.

\subsection{Neutral meson mixing} 
\label{app-NMM}

We calculate the coefficients $P_a^{B_s}(\mu_Z)$, 
defined in Eq.~\eqref{eq-defPa}, 
to evaluate the new physics contribution to the $B_s$-$\ol{B}_s$ mixing.

We refer to the lattice result of the HPQCD collaboration 
for the Bag parameters~\cite{Dowdall:2019bea}, 
\begin{align}
& B_{B_s}^{(1)} = 0.813~(35), 
\quad 
 B_{B_s}^{(2)} = 0.817~(43), 
\quad 
 B_{B_s}^{(3)} = 0.816~(57), 
\notag  \\ 
& B_{B_s}^{(4)} = 1.033~(47), 
\quad 
 B_{B_s}^{(5)} = 0.941~(38),  
\end{align} 
which are evaluated at $\mu_B := m_b(m_b) = 4.162~\GeV$. 
These are mostly consistent with the ETM~\cite{ETM:2013jap} and 
FNAL/MILK~\cite{FermilabLattice:2016ipl} collaborations~\cite{Tsang:2023nay}.  
The values of the matrix elements of the operators in Eq.~\eqref{eq-OBBbar} 
are given by 
\begin{align}
\vev{Q^{\VLL}_1}(\mu_B) =&\ \frac{1}{4} f_{B_s}^2 m_{B_s}^2 \eta_1 B_{B_s}^{(1)},  
\\ \notag 
\vev{Q^{\LR}_1}(\mu_B) =&\ -\frac{1}{2} f_{B_s}^2 m_{B_s}^2 \eta_5 B_{B_s}^{(5)},  
&\quad &
\vev{Q^{\LR}_2}(\mu_B) = \frac{1}{4} f_{B_s}^2 m_{B_s}^2 \eta_4 B_{B_s}^{(4)},  
\\ \notag 
\vev{Q^{\SLL}_1}(\mu_B) =&\ \frac{1}{4} f_{B_s}^2 m_{B_s}^2 \eta_2 B_{B_s}^{(2)},  
&\quad &
\vev{Q^{\SLL}_2}(\mu_B) = f_{B_s}^2 m_{B_s}^2 \left(
                    \eta_2 B_{B_s}^{(2)} + 2\eta_3 B_{B_s}^{(3)}\right).   
\end{align}
Here, 
the coefficients $\eta_i$, with $i=1,2,3,4,5$ and evaluated at $\mu_B$,   
are given by~\cite{Dowdall:2019bea}  
\begin{align}
 \eta_1 =&\ 2.667, 
\quad 
 \eta_2 =  -2.669\;(62), 
\quad 
 \eta_3 = 0.534\;(12), 
\quad 
 \eta_4 = 3.536\;(74), 
\quad 
 \eta_5 = 2.068\;(25).  
\end{align}
The $B_s$ meson mass and decay constant are respectively set at 
$m_{B_s} = 5.36691\;(11)~\GeV$~\cite{ParticleDataGroup:2024cfk} 
and $f_{B_s} = 230.0\;(1.3)~\MeV$~\cite{FlavourLatticeAveragingGroupFLAG:2024oxs}.

The anomalous dimension of the Wilson coefficients at the leading order 
is given by~\cite{Buras:2000if} 
\begin{align}
 \gamma_{M} = 
\begin{pmatrix}
 4 & 0 & 0 & 0  &  0  \\ 
 0 & 2 & 12& 0  &  0  \\
 0 & 0 &-16& 0  &  0  \\ 
 0 & 0 & 0 &-10 &  1/6 \\   
 0 & 0 & 0 &-40 & 34/3 \\   
\end{pmatrix}. 
\end{align}
The evolution matrix can be calculated in the same way as $\ol{B}\to X_s \gamma$. 
Altogether, the coefficients are given by 
\begin{align}
 P_1^{\VLL} =&\ 0.706, 
\quad  
 P_1^{\LR} = -2.25,
\quad  
 P_2^{\LR} = 2.60, 
\quad 
 P_1^{\SLL} = -1.21, 
\quad 
 P_2^{\SLL} = -2.57,   
\end{align}
in the unit of $\GeV^4$ at $\mu = m_Z$.

\section{\texorpdfstring{$\gamma$-$Z$}{gamma-Z} mixing at 1-loop}    
\label{app-AZmixing}

We calculate the 1-loop $\gamma$-$Z$ mixing 
to see the cancellation of the first term in Eq.~\eqref{eq-MaW}.  
The 1-loop corrections for the mixing up to $\order{q^2}$ are given by 
\begin{align}
 \Pi_{\gamma Z}^{\mu\nu}(q^2) =&\  
 \frac{eg}{16\pi^2c_W} \left[ 
 \left(\frac{2}{\eta_W} m_W^2 + \frac{q^2}{3}\right) g^{\mu\nu} 
+\left\{
 \left(\frac{1}{6}+3c_W^2\right)\frac{1}{\eta_W} + \frac{2}{3}c_W^2 
 \right\} \left(q^2g^{\mu\nu} - q^\mu q^\nu \right) 
 \right], 
\notag  \\ 
 \Pi_{\gamma a_0}^\mu(q^2) =&\ 
 \frac{iegm_W}{16\pi^2} 
 \frac{2}{\eta_W} 
  q^\mu,  
\end{align}
where $\Pi_{\gamma Z}^{\mu\nu}$ ($\Pi_{\gamma {a_0}}^\mu$) 
are the mixing between the photon and $Z$ ($a_0$). 
For the mixing with the $Z$ boson, we added 
the diagrams mediated by the $W$ boson, 
its NG modes $a_\pm$ and ghosts $c_\pm$.  
It is straightforward to derive Eq.~\eqref{eq-dltMbsg} 
by using the relation 
\begin{align}
 U^{\dag}_{d_X} H_X U_{d_X} 
= 
 U^{\dag}_{d_X} H_X U_{u_X} U_{u_X}^\dag H_X U_{d_X} 
= 
 \frac{2}{g^2} \left(g^W_X\right)^\dag g^W_X, 
\end{align}
with $X=L,R$.

{\small
\bibliography{ref} 
\bibliographystyle{JHEP} 
}

\end{document}